\documentclass[letterpaper,twocolumn,10pt]{article}
\usepackage{usenix-2020-09}
\usepackage[english]{babel} 
\usepackage[utf8]{inputenc}
\usepackage{blindtext}


\usepackage[all]{nowidow}

\def\BibTeX{{\rm B\kern-.05em{\sc i\kern-.025em b}\kern-.08em
    T\kern-.1667em\lower.7ex\hbox{E}\kern-.125emX}}
    
\usepackage[most]{tcolorbox}
\usepackage{tikz}

\lstdefinestyle{mystyle}{
  commentstyle=\color{codegreen},
  keywordstyle=\color{magenta},
  stringstyle=\color{codepurple},
  basicstyle=\ttfamily\scriptsize,
  breakatwhitespace=false,
  breaklines=true,
  captionpos=b,
  keepspaces=true,
  showspaces=false,
  showstringspaces=false,
  showtabs=false,
  tabsize=2
}

\usepackage[normalem]{ulem}

\usepackage{xcolor,colortbl}

\renewcommand{\sout}[1]{\unskip}

\usepackage{xargs}
\usepackage[colorinlistoftodos,prependcaption,textsize=normalsize]{todonotes}
\usepackage{cleveref}

\newboolean{COMMENTSON} 
\setboolean{COMMENTSON}{true}   
\ifthenelse{\boolean{COMMENTSON}}
{

}

\usepackage[noend,ruled,linesnumbered]{algorithm2e}
\usepackage{booktabs}
\usepackage{listings}
\usepackage[skip=6pt]{caption}

\newcommand{\tech}{\mbox{\textsc{VSP}}}

\usepackage{diagbox}
\usepackage{pifont}

\usepackage{multirow}

\usepackage[explicit,noindentafter]{titlesec}
\titlespacing\section{0pt}{12pt plus 4pt minus 2pt}{8pt plus 2pt minus 2pt}
\titlespacing\subsection{0pt}{12pt plus 4pt minus 2pt}{2pt plus 2pt minus 2pt}
\titlespacing\subsubsection{0pt}{12pt plus 4pt minus 2pt}{0pt plus 2pt minus 2pt}

\usepackage{tikz}
\usepackage{amsmath}

\usepackage[toc,page]{appendix}
\usepackage{filecontents}
\usepackage{algorithmic}
\usepackage{graphicx}
\usepackage{textcomp}

\usepackage{xcolor}
\usepackage[T1]{fontenc} 
\usepackage{mathtools,nccmath}
\usepackage{algorithmic}
\usepackage{graphicx}
\usepackage{textcomp}
\usepackage{xcolor}
\def\BibTeX{{\rm B\kern-.05em{\sc i\kern-.025em b}\kern-.08em
    T\kern-.1667em\lower.7ex\hbox{E}\kern-.125emX}}

\usepackage{hyperref}
\usepackage{color,xcolor}
\usepackage{multirow}
\usepackage{colortbl}
\usepackage{booktabs} 
\usepackage{comment}
\usepackage{tablefootnote}
\usepackage{ifthen}
\usepackage{url}
\usepackage{longtable}
\usepackage{threeparttable}
\usepackage{hyphenat}
\usepackage{bbding}
\usepackage{xspace}
\usepackage[T1]{fontenc}
\usepackage[ruled,linesnumbered]{algorithm2e}
\usepackage{array,multirow,graphicx}
\usepackage{float}
\usepackage{balance}
\usepackage{tikz}
\usepackage{calc}
\usepackage{subfigure}
\usepackage{listings}
\usepackage{pifont}
\usepackage{url}
\usepackage{balance}
\usepackage{xspace}
\usepackage{hyperref,epsfig,endnotes}
\usepackage{array}
\usepackage{paralist}
\usepackage{multirow,makecell}
\usepackage[normalem]{ulem}
\usepackage{pgfplots}
\usepackage{tcolorbox}
\usepackage{courier}
\usepackage{enumitem}
\usepackage{soul}
\usepackage{comment}
\usepackage{url}

\begin{document}

\date{}


\title{Chain-of-Thought Prompting of Large Language Models for Discovering and Fixing Software Vulnerabilities}

\author{
{\rm Yu Nong}\\
Washington State University\\
yu.nong@wsu.edu
\and
{\rm Mohammed Aldeen}\\
Clemson University\\
mshujaa@g.clemson.edu
\and
{\rm Long Cheng}\\
Clemson University\\
lcheng2@clemson.edu 
\and
{\rm Hongxin Hu}\\
University at Buffalo\\
hongxinh@buffalo.edu \\ 
\and
{\rm Feng Chen}\\
The University of Texas at Dallas\\
feng.chen@utdallas.edu \\ 
\and
{\rm Haipeng Cai}\\
Washington State University\\
haipeng.cai@wsu.edu \\ 
} 

\maketitle


\begin{abstract}
Security vulnerabilities are increasingly prevalent in modern software and they are widely consequential to our society. Various approaches to defending against these vulnerabilities have been proposed, among which those leveraging deep learning (DL) avoid major barriers with other techniques hence attracting more attention in recent years. However, DL-based approaches face critical challenges including the lack of sizable and quality-labeled task-specific datasets and their inability to generalize well to unseen, real-world scenarios. 
Lately, large language models (LLMs) have demonstrated impressive potential in various domains by overcoming those challenges, especially through chain-of-thought (CoT) prompting. 
In this paper, we explore how to leverage LLMs and CoT to address three key software vulnerability analysis tasks: identifying a given type of vulnerabilities, discovering vulnerabilities of any type, and patching 
detected vulnerabilities. 
We instantiate the general CoT methodology in the context of these tasks through {\tech}, our unified, vulnerability-semantics-guided prompting approach, and conduct extensive experiments assessing {\tech} versus five baselines 
for the three tasks against three LLMs and two datasets. 
Results show substantial superiority of our CoT-inspired prompting 
(553.3\%, 36.5\%, and 30.8\% higher F1 accuracy for vulnerability identification, discovery, and patching, respectively, on CVE datasets) 
over the baselines. 
Through in-depth case studies analyzing {\tech} 
failures, we also reveal current gaps in LLM/CoT for challenging 
vulnerability cases, while proposing and validating respective improvements. 
\end{abstract}

\section{Introduction}\label{sec:intro}

Software vulnerabilities are consequential~\cite{vulconsequence231}, posing significant threats to security of the cyberspace: they often result in critical financial losses, service disruptions, and data breaches~\cite{vulconsequence232}. The costs associated with remediating these vulnerabilities (e.g., expenses for incident response and system repair) are also substantial~\cite{vulcost23}. 
Meanwhile, security vulnerabilities are pervasive in modern software: according to the NVD statistics~\cite{cvedashboard23}, there have been 
22,378 vulnerabilities publicly reported so far within 2023 alone, and this number has been on a steady rise over the years. In fact, there are even more than reported since some are silently patched~\cite{mssilentpatch23} many have not been discovered yet. 
On the other hand, it is difficult to avoid introducing vulnerabilities during software development and evolution~\cite{iannone2022secret}. Thus, it is crucial to defend against vulnerabilities through vulnerability detection and patching. 

Various approaches to defensive software vulnerability analysis have been proposed, including those based on static/dynamic program analysis~\cite{li2010comparative,austin2013comparison}, the mixture of both~\cite{nong2021evaluating}, and data-driven methods (especially machine/deep learning)~\cite{chakraborty2021deep,nong2022open}. 
These techniques demonstrated respective merits while covering 
all the major kinds of defense tasks against vulnerabilities, including 
\textit{identification} of given types of vulnerabilities~\cite{antunes2009comparing} (e.g., memory leaks~\cite{li2020pca,emamdoost2021detecting}, use-after-free~\cite{nguyen2020binary,wang2020typestate,caballero2012undangle,wu2018fuze}, double-free~\cite{caballero2012undangle}, XSS~\cite{wang2018tt}, and buffer overflow~\cite{haller2013dowser,lhee2002type,ruwase2004practical}), 
\textit{discovery} of vulnerabilities of various types~\cite{kroening2014cbmc,hay2015dynamic,bruening2011practical,nethercote2007valgrind,serebryany2012addresssanitizer,fu2021flowdist,li2022polycruise,wu2022vulcnn,zhou2019devign,mirskyvulchecker,fu2022linevul,hin2022linevd,li2021sysevr}, 
and 
\textit{patching} the detected vulnerabilities~\cite{ma2017vurle,zhang2022example,gao2021beyond,chen2022neural,fu2022vulrepair,pearce2023examining}. 

However, these classes of techniques each face their major challenges. 
In particular, \textit{purely static techniques} (e.g.,~\cite{li2020pca,emamdoost2021detecting,kroening2014cbmc})
suffer from the excessive imprecision of the underlying static analysis~\cite{nong2021evaluating}, causing high rates of false alarms which are a key practicality barrier~\cite{johnson2013don}. 
In addition, they are subject to unsoundness due to the prevalent use of dynamic language constructs~\cite{li2022polycruise} 
in modern programs and limited scalability to large-scale codebases~\cite{li2010comparative,nong2021evaluating}. 
\textit{Purely dynamic} (e.g.,~\cite{bruening2011practical,nethercote2007valgrind,hay2015dynamic}
and \textit{hybrid} (e.g.,~\cite{serebryany2012addresssanitizer}) 
\textit{techniques} have their capabilities constrained by the coverage of existing program inputs, which is usually insufficient, hence missing 
potentially critical vulnerabilities (i.e., low recall~\cite{nong2021evaluating}). 

Data-driven approaches, especially those based on deep learning (DL), helped with overcoming those limitations~\cite{chakraborty2021deep}. 
Indeed, existing \textit{DL-based techniques}, leveraging various neural network architectures (e.g., CNN~\cite{wu2022vulcnn}, RNN/Transformer~\cite{fu2022linevul,hin2022linevd}, and GNN~\cite{zhou2019devign,mirskyvulchecker}), 
have achieved notable successes~\cite{nong2022open}. 
Yet despite their merits in both detection~\cite{mirskyvulchecker,li2021vulnerability,li2021sysevr} and repair~\cite{chen2022neural,fu2022vulrepair}, they fall short of practical performance when applied to unseen, real-world programs~\cite{nong2022open} and stumble on the lack of sizable and quality training datasets~\cite{nong2022generating,nongvulgen}. 
Most recent applications~\cite{nongvulgen,nongvgx} 
of 7 state-of-the-art DL-based techniques~\cite{zhou2019devign,chakraborty2021deep,fu2022linevul,hin2022linevd,li2021vulnerability,chen2022neural,fu2022vulrepair} show that they achieved no more than 16.43\% F1 accuracy and 
8.55\% top-1 accuracy 
for vulnerability detection and repair, respectively, in  replication settings; 
even after using the most advanced data-augmentation technique focused on these tasks~\cite{nongvgx} to augment the training sets with 15,000+ high-quality samples, those two numbers could only go up to 20.1\% and 
21.05\%, 
respectively. 
%

Lately, pre-trained large language models (LLMs) have demonstrated promising performance in assisting with a wide range of tasks, including those of software analysis~\cite{kang2023large,joshi2023repair}. Especially, unlike fine-tuning~\cite{yang2021few} and prompt learning~\cite{liu2023pre}, 
prompting LLMs does not need sizable task-specific datasets, hence addressing the key pain point with the DL-based approaches. In particular, chain-of-thought (CoT)~\cite{wei2022chain}, a state-of-the-art prompting technique has shown 
excitingly potential in pushing up the performance of various LLMs. 
Through, a few step-by-step reasoning demonstrations provided to LLMs as exemplars during prompting, CoT elicits the same reasoning capabilities of the model toward correct answers with a much greater chance. 
While CoT has been explored in relevant tasks~\cite{feng2023prompting,xia2023universal,li2023hitchhiker} and its core idea is simple, it remains unclear how it may be effectively instantiated and leveraged for software vulnerability analysis tasks. 
For instance, we experimented with exemplifying line-by-line code semantics description as the chain of thoughts, but only to no avail. 
On the other hand, using LLMs without reasoning-eliciting prompting (e.g., in the zero-shot setting) has not quite succeeded in vulnerability analysis~\cite{pearce2023examining,wu2023effective}.

In this paper, we systematically explore defensive software vulnerability analysis by prompting LLMs, instantiating the general CoT methodology in the context of vulnerability identification, discovery, and patching tasks. 
Underlying these three tasks, we propose \textit{\textbf{v}ulnerability \textbf{s}emantics guided \textbf{p}rompting} ({\tech}), 
a unified CoT-inspired prompting strategy that maps \textit{vulnerability semantics} (i.e., behaviors of a vulnerable program that make it vulnerable) 
to chains of thoughts. 
With full code semantics being represented by all conventional data and control flow dependencies~\cite{aho2006compilers} in a program, vulnerability semantics are represented by the subset of those dependencies that capture (i.e., account for) the vulnerable behaviors. 
Accordingly, {\tech} formulates the \textit{data/control flow facts underlying vulnerability semantics} as the ``thoughts'' and \textit{the corresponding flow paths} as the ``chain'' in CoT. 
On top of {\tech} as the unified approach, we further propose a specific prompting scheme for each individual task; but all these schemes are guided by vulnerability semantics.

We then conducted extensive experiments, examining the efficacy of {\tech} versus five baselines, including two variants of {\tech}, the full-code-semantics-based prompting strategy (i.e., naive instantiation of CoT), few-shot learning, and standard prompting. 
With these six strategies against three LLMs and two datasets, we compare the different ways of directly leveraging or eliciting the (existing) knowledge of various LLMs for the three vulnerability analysis tasks. 
Our results show that, with just a few (20) exemplars, {\tech} outperformed the baseline approaches by substantial margins for any of the three tasks, especially on the much more challenging dataset. 
For instance, for \textit{vulnerability identification}, our method achieved 65.29\% F1 on a synthetic dataset and \textbf{58.48\%} F1 on a real-world (CVE) dataset, versus the best (non-{\tech}) baseline achieving only 56.28\% and \textbf{8.96\%}, respectively, on GPT-3.5~\cite{floridi2020gpt}. 
For \textit{vulnerability discovery}, {\tech} achieved 54.07\% F1 on the synthetic dataset and \textbf{45.25\%} F1 on the CVE dataset, versus the best baseline achieving 52.04\% and \textbf{33.16\%}, respectively, on Llama2~\cite{touvron2023llama}. 
For \textit{vulnerability patching}, these numbers are 97.65\% and \textbf{20.00\%} with {\tech} versus 96.47\% and \textbf{15.29\%} with the best baseline, also on GPT-3.5. 
{\tech} also helped discover \textbf{22} true zero-day CVE vulnerabilities (with 40.00\% accuracy), versus the standard-prompting baseline found \textbf{9} (with 16.36\% accuracy) only. 

Moreover, we performed in-depth case studies to analyze the failures 
encountered by {\tech}, and 
identified four common root causes across the three tasks. 
For instance, for both the vulnerability identification and patching tasks, and underlying both the false positives and false negatives, 
the dominating failure cause is that the \textit{code context is insufficient} in the code snippet fed to the LLMs (e.g., lacking information about a self-defined/external function called in the code). 
Another major root cause, which is also the primary one for vulnerability discovery failures, is that the \textit{LLMs miss important control flow facts as part of the vulnerability semantics} during their reasoning. 
In contrast, missing key data flow facts is another main failure cause, but more of a problem for vulnerability discovery than for 
identifying or patching vulnerabilities. 
We further proposed improvements 
to overcome these failures per their causes, and show the efficacy of our recommendations (e.g., making up insufficient context via code comments). 

To the best of our knowledge, our work is the first to systematically examine how to leverage prompting LLMs for all three representative tasks in defensive software vulnerability analysis. Through {\tech}, we also demonstrate 
a promising direction toward those tasks 
by guiding LLMs via vulnerability semantics hence eliciting LLM knowledge to reason about the most essential behaviors of a program that make it vulnerable. 
Our extensive experiments and discussions 
suggest this direction can push the effectiveness of software vulnerability analysis to a new height (over the best DL-based approaches achieved so far). 
We also contribute, via our failure case studies and recommended solutions, to future ways of improving vulnerability analysis using LLMs. 
All of the code and datasets for our study have been made 
available at 
\href{https://figshare.com/s/facae52302b6ddf70758}{\underline{Figshare}}.

\begin{figure*}[t]
\centering
	\includegraphics[width=0.9\linewidth]{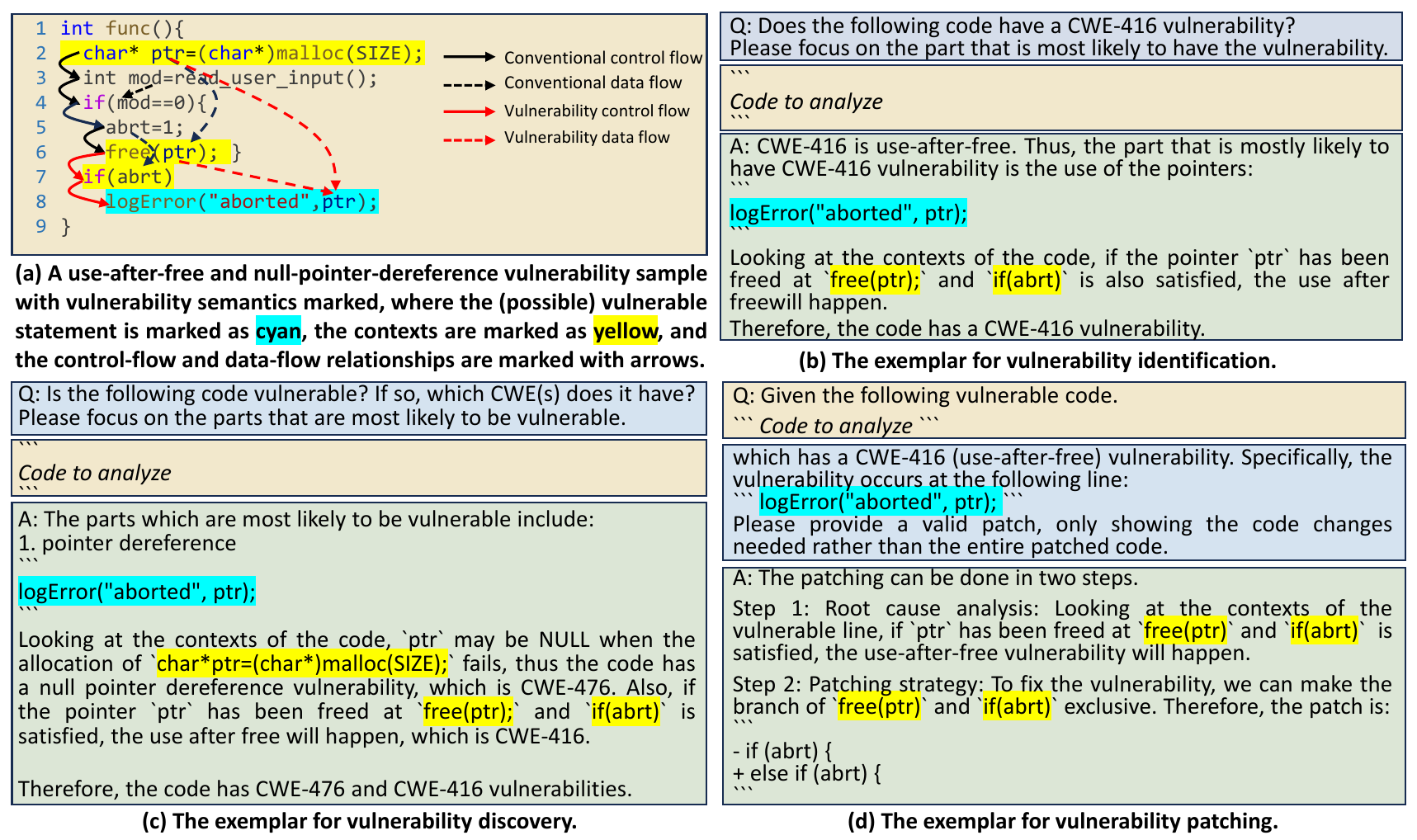}
        \caption{Illustrations of vulnerability semantics (a) and {\tech} exemplars for the three vulnerability analysis tasks (b,c,d).}
	\label{fig:illustrate}
        \vspace{-2pt}
\end{figure*}

\vspace{-2pt}
\section{Preliminary}
\vspace{-2pt}
\textbf{Prompt Engineering.} 
Recent LLMs such as GPT-3.5~\cite{floridi2020gpt} have shown promising performance with various 
tasks~\cite{liu2023your,fan2023large,kang2023large,wei2023copiloting,wang2023software}. Users can directly ask LLMs to solve a problem via a natural language prompt like "whether the following code has a bug". Recent studies~\cite{pearce2023examining,noever2023can} show that the prompting strategies significantly affect the performance of LLMs. To better exploit LLMs, \emph{prompt engineering}~\cite{white2023prompt} which seeks to optimize the prompts becomes a crucial topic. In prompt engineering, few-shot learning~\cite{brown2020language} can also improve performance with a few examples of desired inputs and outputs.

\noindent
\textbf{Chain-of-Thought Prompting.} In prompting engineering with few-shot learning, chain-of-thought (CoT)~\cite{wei2022chain} shows enormous potential on complex reasoning tasks by providing intermediate reasoning steps in the exemplars. It is effective on different tasks such as arithmetic reasoning 
and symbolic reasoning~\cite{wei2022chain}. 
Moreover, zero-shot chain-of-thought is also effective in some tasks by directly showing the reasoning steps in the question without providing exemplars~\cite{kojima2022large}.   

\noindent
\textbf{Vulnerability Analysis using LLMs.}
LLMs have been explored for vulnerability analysis tasks such as vulnerability finding~\cite{noever2023can}, repair~\cite{pearce2023examining}, and secure code generation~\cite{he2023controlling}. However, these works use simple prompts and leave all the vulnerability analysis reasoning process to the LLMs. 
Since those tasks require complex reasoning steps such as vulnerability localization and comprehensive control/data flow analysis, simple prompts may not be effective enough. 
By its nature, CoT seems to fit well here, yet it remains unknown how to best realize its potential for software vulnerability analysis, exploring which is thus the goal of this paper. 

\section{Methodology} \label{sec:meth}
In this section, we introduce our unified prompting strategy for vulnerability analysis. 
Then, we describe the three analysis tasks considered and the 
task-specific prompting schemes all based on the unified strategy. 
Note that our main goal is to explore how to effectively prompt LLMs for vulnerability analysis hence providing evidence about the potential/gaps and actionable insights to inform future technique development, rather than developing a deployable tool here. 
Finally, we describe the datasets and LLMs we use in this study, as well as the design and implementation of our experiments.

\vspace{-3pt}
\subsection{\hspace{-4pt}Vulnerability-Semantics-guided Prompting} \label{sec:strategy}

Inspired by CoT~\cite{wei2022chain}, we propose \emph{Vulnerability-Semantics-guided Prompting (VSP)}, which also instantiates the general CoT methodology. 
This strategy hones in on \emph{vulnerability semantics}, the pivotal components of effective vulnerability analysis. This design is underpinned by two key insights: (1) While a program may encompass numerous lines of code, only a small fraction may be susceptible to vulnerabilities. Thus, centering on these critical segments, constituting the \textit{vulnerability semantics}, empowers LLMs to conduct vulnerability analysis to the fullest extent; (2) Mainstream LLMs are based on Transformer~\cite{vaswani2017attention}, which is known to exhibit diminished performance when handling/generating extensive text~\cite{nong2022generating}. Therefore, concentrating on vulnerability semantics aids in curtailing the length of exemplars, prompts, and answers, mitigating potential distractions for the LLMs.

\begin{figure*}[t]
\centering
	\includegraphics[width=0.8\linewidth]{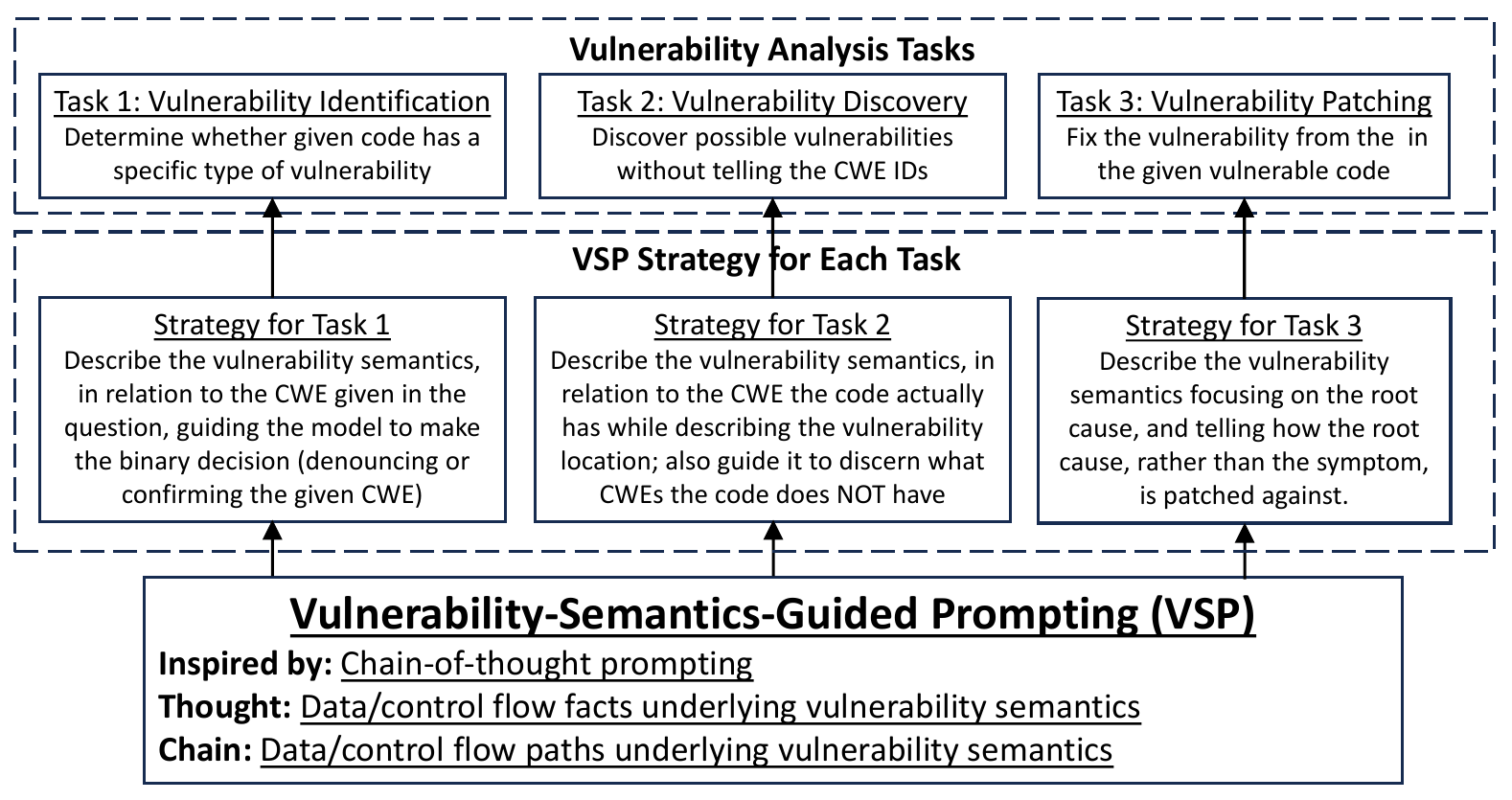}
	\caption{An overview of vulnerability-semantics-guided prompting for vulnerability analysis tasks.}
	\label{fig:overview}
\end{figure*}

\noindent
\textbf{Vulnerability Semantics.} 
Given that only a small proportion of code may cause vulnerabilities, locating the statements that are (possibly) vulnerable is crucial as these statements are the core of vulnerability semantics.
However, such statements themselves cannot cause vulnerabilities without proper context. For a comprehensive vulnerability analysis, the contextual statements that have control and data-flow relationships that may involve \emph{vulnerable behaviors} are also important~\cite{nong2021evaluating,nong2020preliminary}. Therefore, we define \ul{vulnerability semantics} as the behaviors of 
vulnerable statements 
and those of their 
relevant contextual statements. 
This avoids including all the code semantics and helps LLMs focus on the crucial parts.

Figure~\ref{fig:illustrate} (a) shows an example involving use-after-free and null-pointer-dereference vulnerabilities with vulnerability semantics marked. To understand the 
semantics, we first locate the statement that 
likely has 
the two vulnerabilities, which uses a pointer 
at line 8 marked in cyan. Then, we locate the statements that have 
data and control flow dependency relationships with line 8: lines 2 and 6 have the data-flow relationships that may cause a NULL pointer dereference and use-after-free vulnerabilities, as marked with red dashed arrows. Lines 6 and 7 have control-flow relationships that may cause the use-after-free vulnerability, as marked with the red arrows. Therefore, lines 2, 6, and 7 form the context, which are marked in yellow. Therefore, line 8 along with the context constitutes the vulnerability semantics. 

\noindent
\textbf{Prompting Strategy.}
Guided by vulnerability semantics, we leverage CoT-based few-shot learning to achieve {\tech}. 
Each {\tech} prompt consists of two parts. The first part is the set of a few exemplars that provide examples of the reasoning steps for the model to perform few-shot learning. Specifically, each exemplar consists of a question for the task, a code sample, and the answer for the question which includes the reasoning steps. The reasoning steps in the answers focus on the vulnerability semantics discussed above. 
The second part is the testing sample we want the model to analyze. In the testing sample, we ask the question which has the same format as the questions in the exemplars. Then, we expect the LLMs to output the answer following the vulnerability-semantics-based reasoning steps in the exemplars.  

The VSP 
strategy 
lays the bedrock of 
different vulnerability analysis tasks in our study. 
Figure~\ref{fig:overview} 
illustrates the overall design: 
We use {\tech} as the unified/common 
strategy under  
the detailed prompting scheme for each task. Note that we intend to examine the potential of VSP for each task---these tasks are orthogonal, and each task assumes what it needs to assume, rather than based on the output of the previous task. We do not aim to develop a holistic vulnerability analysis system (e.g., starting with identification followed by patching).

\subsection{Tasks}
In this work, we address three software vulnerability analysis tasks: (1) vulnerability \textbf{identification} (determining if a provided code sample has a specific type of vulnerability), (2) vulnerability \textbf{discovery} (discovering potential vulnerabilities in code samples without specifying their types), and (3) vulnerability \textbf{patching} (creating a patch based on a vulnerable sample and vulnerability type and location). 
These tasks span various levels of vulnerability analysis, representing the key steps in defenses against software vulnerabilities.

Next, we separately describe the goals, formulation, and detailed prompting schemes for the three tasks.

\subsubsection{Task 1: Vulnerability Identification}
\textbf{Goal.} The goal of vulnerability identification is to determine whether a given code sample has a specific type of vulnerability (e.g., CWE), as whether a vulnerability detector can find specific types of vulnerabilities determines its effectiveness.

\noindent
\textbf{Formulation.} We formulate vulnerability identification as a binary classification task. In the LLM input prompt, we provide a code sample in terms of text and ask the model \emph{whether the given text has a CWE-xxx vulnerability}, where \emph{CWE-xxx} is the CWE ID of the vulnerability type. 


\noindent
\textbf{Prompting.} As discussed in Section~\ref{sec:strategy}, we provide the exemplars for vulnerability identification and the testing sample in each prompt. As Figure~\ref{fig:illustrate} (b) shows, for each exemplar, we ask the question for vulnerability identification: "\textit{Q: Does the following code have a CWE-xxx vulnerability?}".  Besides the question itself, we also request the model: "Please focus on the part that is most likely to have the vulnerability." This is to guide the model to focus on vulnerability semantics. 
Then, we provide the code to analyze. 
Each exemplar's answer part includes an explanation of the CWE ID's meaning for clarity. 
Then, we find the possible vulnerable statements based on the vulnerability type as well as the contexts of them based on the vulnerability control and data flow relationships. We analyze the conditions of the contexts and decide whether the code has a specific type of vulnerability. We finally provide the testing sample by asking the same question as in the exemplars to let the model output its response.


\subsubsection{Task 2: Vulnerability Discovery}
\textbf{Goal.} 
%
In this task, LLMs are asked to discover possible vulnerabilities without being provided with CWE IDs. This serves to evaluate their practical applicability. In real-world software vulnerability analysis, developers often lack prior knowledge of specific vulnerabilities in given code and their corresponding CWE IDs. Hence, vulnerability analysis techniques need to pinpoint potential vulnerabilities and classify their types.

\noindent
\textbf{Formulation.}
We formulate vulnerability discovery as a multi-class classification task. Given a code sample in the form of text, we ask the model whether the code has any vulnerability(ies), and if so, which CWE(s) the code has. 

\noindent
\textbf{Prompting.} 
As Figure~\ref{fig:illustrate} (c) shows, in each exemplar, we ask the question for vulnerability discovery: "\textit{Q: Is the following code vulnerable? If so, which CWE(s) does it have? Please focus on the parts that are most likely to be vulnerable.}" Then, we provide the code sample. 
In the answer, we enumerate the types of statements that may be vulnerable and locate them within the code. For each potentially vulnerable statement, we identify the associated contexts based on the relevant vulnerability control and data flow relationships. Ultimately, we conduct an analysis of the vulnerability semantics to determine the presence of vulnerability(-ies) and, if any, their corresponding CWE(s). Subsequently, we present the test sample with the same question as posed in the exemplars, letting the model provide its response.



\subsubsection{Task 3: Vulnerability Patching} 
\textbf{Goal.} The goal of this task is to eliminate the vulnerability from the code while preserving the code functionality. This is crucial because it addresses the security issues of the code and avoids possible attacks and damage to the systems. 

\noindent
\textbf{Formulation.}
We formulate vulnerability patching as a text-to-text transformation task. Given a vulnerable sample in the form of text, we ask the model to generate a patch. 
We directly tell the model the vulnerable line and the 
respective CWE, 
which 
is common for bug/vulnerability patching~\cite{fu2022vulrepair,chen2022neural,gao2021beyond}. With the vulnerable code sample, the vulnerable line, and the CWE ID, the model outputs the patch 
in terms of only 
the lines changed (rather than the entire patched sample). 

\noindent
\textbf{Prompting.}
As Figure~\ref{fig:illustrate} (d) shows, in each exemplar, we ask the model to generate a patch for the given code: "\textit{Given the following vulnerable code: ```<vulnerable code>``` which has a CWE-xxx (<description of CWE>) vulnerability. Specifically, the vulnerability occurs at the following line: <vulnerable line>. Please provide a valid patch, only showing the code changes needed rather than the entire patched code.}" 
In the answer, we 
analyze the \emph{root cause} of the vulnerability based on its semantics. Then, 
we assess the \emph{patching strategy} and identify a corrective action to fix the vulnerability via code edits. Finally, we provide a test sample with the same question as in the exemplars, letting the model give its response.


\subsection{Datasets}
%
%
%
In this study, we manually construct exemplars for CoT-based few-shot learning, targeting five most dangerous CWEs in C/C++ code~\cite{cwereport}: (1) CWE-787 (out-of-bound write), (2) CWE-125 (out-of-bound read), (3) CWE-476 (NULL-pointer-dereference), (4) CWE-416 (use-after-free), and (5) CWE-190 (integer overflow). 
We focus on C/C++ because C/C++ vulnerabilities cover the dominant classes of vulnerabilities and they are the most vulnerable languages~\cite{nong2021evaluating}. 
For each CWE, we provide four pairs of samples, each comprising a vulnerable and a corresponding patched sample. This results in a total of 20 pairs. The samples are inspired by the examples on the CWE report website~\cite{cwereport}. We use the strategies described above to write the respective exemplars.

%
%

We evaluate the three tasks using two testing datasets. The first is a synthetic dataset known as SARD~\cite{black2017sard}, containing a substantial number of vulnerable and respective non-vulnerable samples. The second is a real-world vulnerability dataset curated by Fan et al.\cite{fan2020ac}, sourced from CVE reports\cite{nvd}, providing both vulnerable and respective patched code samples. We exclusively focus on samples pertaining to the five CWEs mentioned above. Considering the cost of commercial LLMs (e.g., GPT-3.5) and the time consumption, we only select 500 pairs of samples for each testing dataset. 
We use both datasets as we want to assess the difficulties of analyzing simple/synthetic samples and complex/real-world samples with LLMs, helping us obtain more in-depth insights.

\subsection{Models}




Table~\ref{tab:llm} summarizes the three latest LLMs used in our study, including the model version, size, and token constraints. 
In particular, \textbf{GPT-3.5}~\cite{floridi2020gpt} is a primary LLM from OpenAI which can understand and generate natural language and code. 
We chose GPT-3.5-Turbo over GPT-4 due to the latter's high price. 
\textbf{Llama2}~\cite{touvron2023llama} is the latest LLM from Meta AI. 
We chose the {\tt llama-2-7b-chat-hf} version due to the resource limitation of our machine. 
For the same reason, we chose the {\tt falcon-7b-instruct} version of \textbf{Falcon}~\cite{zxhang2023falcon}, an LLM from technology innovation institute.
While these models are not the most advanced/capable in their families, our choices 
should not affect much what our study aims to achieve. 


\begin{table}[t]
    \centering
    \scalebox{0.8}{
    \begin{tabular}{|c|c|c|c|}
    \hline
    Model&Version  & \#Param& \#Max Tokens \\ \hline
    GPT-3.5 & gpt-3.5-turbo-16k & 175B & 16,385 \\ 
    Llama2 & llama-2-7b-chat-hf & 7B & 4,096 \\
    Falcon & falcon-7b-instruct & 7B & 2,048 \\
    \hline
    \end{tabular}}
    \caption{LLMs we investigated.}
    \label{tab:llm}
    \vspace{-4pt}
\end{table}


\vspace{-2pt}
\subsection{Baselines}
In addition to {\tech}, we examine the following five baselines:

\textbf{Standard Prompting.} We 
ask the LLMs to address the tasks without providing any exemplars or guidance, aiming to assess their raw/original potential. 

\textbf{Standard Few-Shot Learning.} For each exemplar, we only provide the final answer without any reasoning steps. This is to test whether the improvements (if any) are from the reasoning steps rather than the final answers in the exemplars.

\textbf{Naive CoT-based Few-Shot Learning.} Instead of focusing on the vulnerability semantics, we analyze the code line by line to address the tasks. This is to test whether the VSP strategy outperforms full semantics analysis.  

\textbf{Zero-Shot VSP Prompting.} We do not provide the exemplars, but in the prompts, we provide steps for VSP-based vulnerability analysis. This is to test whether providing exemplars improves the model's reasoning capabilties. 

\textbf{Other-type VSP Prompting.} 
We ensure that the exemplars used do not overlap with any testing sample in terms of the CWE covered, for assessing whether LLMs' reasoning capabilities elicited from the 
exemplars can be transferred across different vulnerability types.


\vspace{-2pt}
\subsection{Implementation}
For GPT-3.5, we subscribe the APIs from OpenAI to process the input and generate the results. For Llama2 and Falcon, we download the models from HuggingFace~\cite{wolf2019huggingface} and use its APIs to conduct the experiments. We set the temperatures of the LLMs to 0 to reduce the randomness. We perform our experiments on a workstation with an AMD Ryzen Threadripper Pro 5595WX (4.5GHz) CPU with 64 Cores, four Nvidia GeForce RTX A6000 GPU, and 512GB memory.

\begin{table*}[t]
  \centering
  \scalebox{0.7}{
\begin{tabular}{|c|c|l|r|c|c|l|r|c|c|l|r|}
\hline
Model & Dataset & \multicolumn{1}{c|}{Strategy} & \multicolumn{1}{c|}{F1} & Model & Dataset & \multicolumn{1}{c|}{Strategy} & \multicolumn{1}{c|}{F1} & Model & Dataset & \multicolumn{1}{c|}{Strategy} & \multicolumn{1}{c|}{F1} \\
\hline
\multirow{12}[4]{*}{GPT-3.5} & \multirow{6}[2]{*}{SARD} & Standard Prompting & \cellcolor[rgb]{ .475,  .78,  .557}56.28\% & \multirow{12}[4]{*}{Llama2} & \multirow{6}[2]{*}{SARD} & Standard Prompting & \cellcolor[rgb]{ .486,  .784,  .565}54.95\% & \multirow{12}[4]{*}{Falcon} & \multirow{6}[2]{*}{SARD} & Standard Prompting & \cellcolor[rgb]{ .988,  .988,  1}0.00\% \\
      &       & Standard Few-Shot & \cellcolor[rgb]{ .678,  .863,  .733}34.01\% &       &       & Standard Few-Shot & \cellcolor[rgb]{ .471,  .78,  .553}56.71\% &       &       & Standard Few-Shot & \cellcolor[rgb]{ .529,  .804,  .604}50.15\% \\
      &       & Naive CoT Learning & \cellcolor[rgb]{ .851,  .933,  .882}15.17\% &       &       & Naive CoT Learning & \cellcolor[rgb]{ .553,  .812,  .624}47.64\% &       &       & Naive CoT Learning & \cellcolor[rgb]{ .78,  .906,  .824}22.69\% \\
      &       & Zero-Shot VSP & \cellcolor[rgb]{ .478,  .784,  .561}55.78\% &       &       & Zero-Shot VSP & \cellcolor[rgb]{ .502,  .792,  .58}53.25\% &       &       & Zero-Shot VSP & \cellcolor[rgb]{ .98,  .984,  .992}1.19\% \\
      &       & VSP Prompting & \cellcolor[rgb]{ .388,  .745,  .482}\textbf{65.29\%} &       &       & VSP Prompting & \cellcolor[rgb]{ .435,  .765,  .525}\textbf{60.18\%} &       &       & VSP Prompting & \cellcolor[rgb]{ .518,  .8,  .596}\textbf{51.25\%} \\
      &       & Other-Type VSP & \cellcolor[rgb]{ .392,  .749,  .486}65.06\% &       &       & Other-Type VSP & \cellcolor[rgb]{ .443,  .769,  .533}59.33\% &       &       & Other-Type VSP & \cellcolor[rgb]{ .525,  .804,  .6}50.46\% \\
\cline{2-4}\cline{6-6}\cline{10-10}      & \multirow{6}[2]{*}{CVE} & Standard Prompting & \cellcolor[rgb]{ .957,  .976,  .973}3.59\% &       & \multirow{6}[2]{*}{CVE} & Standard Prompting & \cellcolor[rgb]{ .667,  .859,  .725}35.10\% &       & \multirow{6}[2]{*}{CVE} & Standard Prompting & \cellcolor[rgb]{ .949,  .973,  .965}4.69\% \\
      &       & Standard Few-Shot & \cellcolor[rgb]{ .984,  .988,  .996}0.80\% &       &       & Standard Few-Shot & \cellcolor[rgb]{ .678,  .863,  .733}34.08\% &       &       & Standard Few-Shot & \cellcolor[rgb]{ .71,  .878,  .761}30.45\% \\
      &       & Naive CoT Learning & \cellcolor[rgb]{ .91,  .957,  .929}8.96\% &       &       & Naive CoT Learning & \cellcolor[rgb]{ .78,  .906,  .824}22.72\% &       &       & Naive CoT Learning & \cellcolor[rgb]{ .776,  .902,  .816}23.40\% \\
      &       & Zero-Shot VSP & \cellcolor[rgb]{ .949,  .973,  .965}4.62\% &       &       & Zero-Shot VSP & \cellcolor[rgb]{ .62,  .839,  .682}40.15\% &       &       & Zero-Shot VSP & \cellcolor[rgb]{ .769,  .902,  .812}23.91\% \\
      &       & VSP Prompting & \cellcolor[rgb]{ .451,  .773,  .537}\textbf{58.48\%} &       &       & VSP Prompting & \cellcolor[rgb]{ .58,  .824,  .647}\textbf{44.80\%} &       &       & VSP Prompting & \cellcolor[rgb]{ .655,  .855,  .714}\textbf{36.36\%} \\
      &       & Other-Type VSP & \cellcolor[rgb]{ .471,  .78,  .557}56.34\% &       &       & Other-Type VSP & \cellcolor[rgb]{ .584,  .827,  .651}44.02\% &       &       & Other-Type VSP & \cellcolor[rgb]{ .667,  .859,  .722}35.41\% \\
\hline
\end{tabular}%
}
     \caption{Vulnerability Identification Results in terms of F1.}
     \label{tab:iden-f1}
     \vspace{-2pt}
\end{table*}%

\vspace{-2pt}
\section{Task 1: Vulnerability Identification}
\vspace{-2pt}
\subsection{Metrics}
Since the task is formulated as a binary classification task, we evaluate the effectiveness in terms of recall, precision, and F1, which are standard metrics for binary classification tasks. We use an automatic script to check the outputs by some key phases (e.g., "the code has a CWE-xxx vulnerability") to determine the model predictions.

\subsection{Results}
\textbf{Overall Performance.}
Table~\ref{tab:iden-f1} shows the effectiveness of \emph{VSP prompting} and the baselines against the vulnerability identification task in terms of F1, because F1 is the metric that represents the overall effectiveness of binary classification. We show the detailed numbers of recall, precision, and F1 for each CWE in Tables~\ref{tab:iden-recall-full}, \ref{tab:iden-precision-full}, and \ref{tab:iden-f1-full} in Appendix. For all three LLMs, VSP prompting achieves the best F1 scores. On the SARD dataset, GPT-3.5, Llama2, and Falcon achieve 65.29\%, 60.18\%, and 51.25\% F1, respectively. On the CVE dataset, GPT-3.5, Llama2, and Falcon achieve 58.48\%, 44.80\%, and 36.36\% F1, respectively.  We also notice that the F1 scores on the SARD dataset are higher than the ones on the CVE dataset. This is because the samples in SARD are synthetic which is generally simple while the samples in CVE are real-world and complex which makes the vulnerability identification more challenging. The high F1 scores indicate that VSP Prompting is practical for vulnerability identification.

\textbf{Comparison to Standard Prompting.} 
To show that VSP prompting is necessary and effective for vulnerability identification, we compare it with simple standard prompting on LLMs. Table~\ref{tab:iden-f1} shows the F1 scores of standard prompting on the same datasets. We notice that the F1 scores of standard prompting are much lower than the ones of VSP prompting on all the settings: with standard prompting, GPT-3.5, Llama2, and Falcon achieve 56.28\%, 54.94\%, and 0.00\% on the SARD dataset, and 3.59\%, 35.10\%, 4.69\% on the CVE dataset. This indicates that the reasoning steps in VSP prompting help LLMs outperform their own reasoning steps for vulnerability identification. We also notice that the improvements of VSP prompting are higher on the CVE dataset (e.g., 58.48\% to 3.59\% for GPT-3.5). This further indicates the necessity of VSP prompting for real-world vulnerability identification.

\textbf{Comparison to Standard Few-shot Learning.}
Table~\ref{tab:iden-f1} shows the F1 scores of standard few-shot learning for vulnerability identification. We notice that it has poor effectiveness compared with VSP prompting. With standard few-shot learning, GPT-3.5, Llama2, and Falcon only achieve 34.01\%, 56.71\%, and 50.15\% F1 scores on the SARD dataset, and 0.80\%, 34.08\%, and 30.45\% F1 scores on the CVE dataset. This indicates that while providing the same exemplars, the LLMs still cannot achieve the effectiveness of VSP prompting without providing the vulnerability-semantics-guided reasoning steps. This indicates the importance of the vulnerability semantics for vulnerability identification.

\textbf{Comparison to Naive CoT Learning.}
To show that our vulnerability-semantic-guided design in VSP prompting is important, we compare it with a naive chain-of-thought design which analyzes the code line by line to identify the vulnerability. As Table~\ref{tab:iden-f1} shows, with naive CoT learning, GPT-3.5, Llama2, and Falcon only achieve 15.17\%, 47.64\%, and 22.69\% F1 scores on the SARD dataset, and 8.96\%, 22.72\%, and 23.40\% F1 scores on the CVE dataset. These numbers are much lower than the ones for VSP prompting. This indicates the importance of vulnerability semantics. Because only a small proportion of code is indeed relevant to vulnerabilities, the line-by-line reasoning involve much information not relevant to vulnerabilities which distracts the LLMs.

\textbf{Comparison to Zero-Shot VSP.} To show the necessity of exemplars, we test zero-shot VSP, which tells the VSP-based reasoning steps in the prompt but does not provide exemplars. We tell the LLMs to "focus on the parts that are most likely to be vulnerable and their contexts based on control-flow and data-flow relationships". As Table~\ref{tab:iden-f1} shows, with zero-shot VSP, GPT-3.5, Llama2, and Falcon only achieve 55.78\%, 53.25\%, and 1.19\% F1 scores on the SARD dataset, and 4.62\%, 40.15\%, and 23.91\% F1 scores on the CVE dataset. These numbers are much lower than the ones for VSP prompting, indicating the importance of exemplars.  

\textbf{Transferability of VSP Prompting.} 
In the experiments above, the CWEs of the testing samples are consistent with the exemplars. In this experiment, we test whether the benefits of VSP Prompting still exist when the CWEs of exemplars are different from the testing samples. We write exemplars with different CWEs (e.g., CWE-20, CWE-78, CWE-269, CWE-369, and CWE-798) but keep the VSP strategies. As Table~\ref{tab:iden-f1} rows "Other-Type VSP" shows, GPT-3.5, Llama2, and Falcon achieve 65.06\%, 59.33\%, and 50.46\% F1 scores on the SARD dataset, and 56.34\%, 44.02\%, and 35.41\% F1 scores on the CVE dataset. These numbers are comparable to the ones in VSP prompting. This indicates that VSP prompting is transferable for vulnerability identification: the LLMs are able to learn the reasoning steps while the CWEs are different.

\begin{table}
    \centering
    \scalebox{0.7}{
    \begin{tabular}{|c|c|c|c|c|}
    \hline
    &\multicolumn{2}{c|}{CVE}& \multicolumn{2}{c|}{SARD} \\ \hline
    Failure Reason& FN & FP & FN & FP \\ \hline
    Insufficient contexts& 35\% & 42\% & 8\% & 7\% \\ 
    Oblivion of CWE& 19\% & 17\% & 17\% & 9\% \\ 
    Incomplete control flow analysis& 18\% & 21\% & 28\% & 40\% \\ 
    Incomplete data flow analysis& 28\% & 20\% & 47\% & 44\%\\ 
    \hline
    \end{tabular}}
    \caption{Failure Reasons for Vulnerability Identification.}
    \label{tab:iden-failure}
    \vspace{-4pt}
\end{table}

\subsection{Failure Case Studies} \label{ssec:iden-failure}
Although VSP prompting outperforms all the other baselines for vulnerability identification, we conduct case studies for the failure cases. Since the case studies are time-consuming, we only do case studies on the GPT-3.5 results. We manually inspect the false negative and false positive cases on the CVE and SARD datasets separately. For each case study, we follow the error analysis in~\cite{he2023you} and randomly select 100 test samples with wrong predictions, and manually inspect the reasons for the failures. Based on the inspection, we categorize the reasons into 4 categories.  

Table~\ref{tab:iden-failure} shows the statistics of the four failure reasons. The first category is insufficient contexts. Because most LLMs have limitations on the number of tokens processed, it is not possible to directly input the whole project into LLMs~\cite{pearce2023examining}. Therefore, in this study, we only input the vulnerable function itself to the LLMs. However, many vulnerabilities are inter-procedural where the vulnerability cannot be comprehensively analyzed without scanning all the code in the project.
We notice that this is common on the CVE dataset, with 35\% on the FN cases and 42\% on the FP cases. This is because the samples from CVE are from real-world projects and they usually have many functions and files, which make the analysis challenging. In comparison, the proportions on the SARD dataset are much lower, with 8\% on the FN cases and 7\% on the FP cases. This is because the synthetic samples from SARD are relatively simple. Most of the samples are standalone and thus the vulnerability identification is much easier. 

The second category is the oblivion of CWE. We notice that in some failure cases, GPT-3.5 "forgets" the CWE type it needs to identify. For example, we ask the model to identify a CWE-787 out-of-bound vulnerability in the question, but the model identifies the NULL pointer dereference vulnerability in its reasoning steps. A possible reason is that the Transformer architecture, where LLMs are based, would have performance degradation when the text to be processed is too long~\cite{nong2022generating}. 
As Table~\ref{tab:iden-failure} shows, the proportions of oblivion of CWE are 19\% and 17\% for the FN cases and FP cases from the CVE dataset respectively, and 17\% and 9\% for the FN cases and FP cases from the SARD dataset respectively, indicating that this issue is widespread.  

The third category is incomplete control flow analysis. In some failure cases, GPT-3.5 is not able to do a comprehensive control flow analysis. For example, when identifying a CWE-476 NULL-pointer-dereference vulnerability, the model may still affirm that the dereference of a pointer is possible to be NULL, while the NULL pointer checking has been done before the dereference. This indicates that while we have provided the control-flow analysis reasoning steps in the exemplars, 
they still cannot do complex and comprehensive analysis like traditional static analysis techniques do. As Table~\ref{tab:iden-failure} shows, the proportions of incomplete control flow analysis are 18\% and 21\% for the FN cases and FP cases from CVE respectively, and 28\% and 40\% for the FN cases and FP cases from SARD respectively. The higher proportions on SARD indicate that, while the samples are simple, the LLMs still have limitations on doing complete control flow analysis.

The fourth category is incomplete data flow analysis. In some failure cases, GPT-3.5 is not able to do comprehensive data flow analysis. For example, when identifying a CWE-787 out-of-bound vulnerability, the model may still affirm that a buffer write may be beyond the boundary of the buffer while the buffer has reserved enough space for the buffer write. This indicates, again, that while LLMs are capable to some extent, they still cannot do complex and comprehensive data flow analysis. As Table~\ref{tab:iden-failure} shows, the proportions of incomplete data flow analysis are 28\% and 20\% for the FN cases and FP cases from  CVE  respectively, and 47\% and 44\% for the FN cases and FP cases from SARD respectively. 



\section{Task 2: Vulnerability Discovery}
%
%
\subsection{Metrics}
Since the task is formulated as a multi-class classification task where the classes are the CWE IDs of the vulnerabilities, we first evaluate the recall, precision, and F1 on each of the classes and then compute the overall effectiveness using macro-averaging and micro-averaging~\cite{aota2020automation}. We use an automatic script to check the outputs of the models like what we do in Task 1. Once the predicted classes involve the ground-truth one, the prediction is positive, otherwise negative.

\subsection{Results}
\textbf{Overall Performance.} Table~\ref{tab:discovery-f1} shows the effectiveness of VSP prompting and the baselines against the vulnerability discovery task in terms of macro-averaged and micro-averaged F1. We show detailed recall, precision, and F1 for each CWE in Tables \ref{tab:discovery-recall}, \ref{tab:discovery-precision}, and ~\ref{tab:discovery-f1-full} in Appendix. We notice that VSP prompting achieves the best overall effectiveness in most of the cases. On the SARD dataset, GPT-3.5, Llama2, and Falcon achieve 55.83\%, 54.07\%, and 46.01\% macro-averaged F1 with VSP prompting, respectively, and the numbers of micro-averaged F1 are 60.21\%, 60.92\%, and 46.83\% respectively. On the CVE dataset, GPT-3.5, Llama2, and Falcon achieve 36.34\%, 45.25\%, and 46.01\% macro-averaged F1 with VSP prompting, respectively, and the numbers of micro-averaged F1 are 36.89\%, 48.35\%, and 28.11\% respectively. This indicates that our VSP prompting strategy is able to improve the LLMs for vulnerability discovery significantly. 

\textbf{Comparison to Standard Prompting.} We compare our VSP prompting to standard prompting. As Table~\ref{tab:discovery-f1} show. On the SARD dataset, GPT-3.5, Llama2, and Falcon achieve 54.02\%, 16.22\%, and 0.17\% macro-averaged F1 with standard prompting, respectively, and the numbers for micro-averaged F1 are 51.50\%, 48.27\%, and 0.39\% respectively. On the CVE dataset, GPT-3.5, Llama2, and Falcon achieve 3.28\%, 17.09\%, and 0.69\% macro-averaged F1 with standard prompting, respectively, and the numbers for micro-averaged F1 are 6.70\%, 35.39\%, and 1.22\% respectively. These numbers are much less than the ones for VSP prompting, indicating that VSP prompting helps LLMs with vulnerability discovery. 

\textbf{Comparison to Standard Few-Shot Learning.} As Table~\ref{tab:discovery-f1} shows, on SARD, GPT-3.5, Llama2, and Falcon achieve 55.19\%, 21.48\%, and 13.78\% macro-averaged F1 with standard few-shot learning, respectively, and the numbers for micro-averaged F1 are 46.83\%, 40.91\%, and 12.46\% respectively. On CVE, GPT-3.5, Llama2, and Falcon achieve 16.21\%, 24.92\%, and 13.78\% macro-averaged F1 with standard prompting, respectively, and the numbers for micro-averaged F1 are 11.60\%, 28.79\%, and 15.06\% respectively. We notice that, while standard few-shot learning has improvements compared to standard prompting in some cases (e.g., on CVE with the GPT-3.5 model), it is not comparable to the VSP prompting. This indicates that the reasoning steps in our VSP prompting are crucial for vulnerability discovery.

\textbf{Comparison to Naive CoT Learning.} As Table~\ref{tab:discovery-f1} shows, on SARD, GPT-3.5, Llama2, and Falcon achieve 41.13\%, 52.04\%, and 18.81\% macro-averaged F1 with naive CoT learning, respectively, and the numbers for micro-averaged F1 are 34.08\%, 59.47\%, and 22.90\% respectively. On CVE, GPT-3.5, Llama2, and Falcon achieve 14.75\%, 33.16\%, and 12.54\% macro-averaged F1 with standard prompting, respectively, and the numbers for micro-averaged F1 are 14.86\%, 39.17\%, and 11.52\% respectively. While naive CoT learning improves LLMs for vulnerability discovery compared to standard prompting in some cases (e.g., on CVE with GPT-3.5), the effectiveness is still lower than one from VSP prompting. This indicates that VSP prompting which focuses on vulnerability semantics is effective for vulnerability discovery.

\textbf{Comparison to Zero-Shot VSP.} 
For zero-shot VFP, we again tell the model to "focus on the parts that are most likely to be vulnerable and its contexts based on control-flow and data-flow relationships". As Table~\ref{tab:discovery-f1} shows, on SARD, GPT-3.5, Llama2, and Falcon achieve 48.49\%, 15.58\%, and 0.63\% macro-averaged F1 with zero-shot VSP, respectively, and the numbers for micro-averaged F1 are 47.91\%, 47.56\%, and 0.79\% respectively. On CVE, GPT-3.5, Llama2, and Falcon achieve 16.81\%, 9.96\%, and 0.00\% macro-averaged F1 with standard prompting, respectively, and the numbers for micro-averaged F1 are 22.07\%, 23.97\%, and 0.00\% respectively. This indicates that, without exemplars in the prompt, the LLMs cannot perform well for vulnerability discovery.

\textbf{Transferability of VSP Prompting.} We again test the transferability of VSF Prompting with the exemplars where the CWEs are different from the testing samples.
The rows "Other-type VSP" in Table~\ref{tab:discovery-f1} show the F1 scores. We notice that on the GPT-3.5 model, the benefits still exist. It achieves 57.33\% macro-averaged F1 and 62.19\% micro-averaged F1 on SARD, as well as 19.58\% macro-averaged F1 and 30.28\% micro-averaged F1 on CVE. However, on the Llama2 and Falcon models, the benefits do not exist. On SARD, Llama2 and Falcon only achieve 29.54\% and 4.28\% macro-averaged F1, and the numbers for micro-averaged F1 are 48.44\% and 8.13\%. On CVE, Llama2 and Falcon only achieve 9.20\% and 1.15\% macro-averaged F1, and the numbers for micro-averaged F1 are 23.83\% and 3.19\%. A possible reason is that we only use the lite versions with 7B parameters for Llama2 and Falcon because of resource limitations. In comparison, we use the full version of GPT-3.5-Turbo with 175B parameters because it provides online API service. This indicates that the capability of LLMs impacts the effectiveness of the transferability of VSP prompting for vulnerability discovery.

\begin{table}[t]
  \centering
  \scalebox{0.7}{
\begin{tabular}{|c|c|lrr|}
\hline
\multicolumn{1}{|l|}{Model} & \multicolumn{1}{l|}{Dataset} & \multicolumn{1}{l|}{Strategy} & \multicolumn{1}{l}{Macro avg} & \multicolumn{1}{l|}{Micro avg} \\
\hline
\multirow{12}[3]{*}{GPT-3.5} & \multirow{6}[2]{*}{SARD} & Standard Prompt & \cellcolor[rgb]{ .471,  .78,  .553}54.02\% & \cellcolor[rgb]{ .494,  .788,  .573}51.50\% \\
      &       & Standard Few-shot & \cellcolor[rgb]{ .459,  .773,  .541}55.19\% & \cellcolor[rgb]{ .537,  .808,  .612}46.83\% \\
      &       & Naïve CoT Learning & \cellcolor[rgb]{ .592,  .827,  .659}41.13\% & \cellcolor[rgb]{ .663,  .859,  .718}34.08\% \\
      &       & Zero-Shot VSP & \cellcolor[rgb]{ .522,  .8,  .6}48.49\% & \cellcolor[rgb]{ .529,  .804,  .604}47.91\% \\
      &       & VSP Prompting & \cellcolor[rgb]{ .451,  .773,  .537}55.83\% & \cellcolor[rgb]{ .408,  .753,  .502}60.21\% \\
      &       & Other-Type VSP & \cellcolor[rgb]{ .435,  .765,  .525}\textbf{57.33\%} & \cellcolor[rgb]{ .388,  .745,  .482}\textbf{62.19\%} \\
\cline{2-5}      & \multirow{6}[1]{*}{CVE} & Standard Prompt & \cellcolor[rgb]{ .957,  .976,  .976}3.28\% & \cellcolor[rgb]{ .925,  .965,  .945}6.70\% \\
      &       & Standard Few-shot & \cellcolor[rgb]{ .835,  .925,  .867}16.21\% & \cellcolor[rgb]{ .878,  .945,  .906}11.60\% \\
      &       & Naïve CoT Learning & \cellcolor[rgb]{ .847,  .933,  .878}14.75\% & \cellcolor[rgb]{ .847,  .933,  .878}14.86\% \\
      &       & Zero-Shot VSP & \cellcolor[rgb]{ .827,  .925,  .863}16.81\% & \cellcolor[rgb]{ .776,  .902,  .82}22.07\% \\
      &       & VSP Prompting & \cellcolor[rgb]{ .639,  .847,  .698}\textbf{36.34\%} & \cellcolor[rgb]{ .635,  .847,  .694}\textbf{36.89\%} \\
      &       & Other-Type VSP & \cellcolor[rgb]{ .8,  .914,  .839}19.58\% & \cellcolor[rgb]{ .698,  .871,  .749}30.28\% \\ \hline
\multirow{12}[2]{*}{Llama2} & \multirow{6}[1]{*}{SARD} & Standard Prompt & \cellcolor[rgb]{ .835,  .925,  .867}16.22\% & \cellcolor[rgb]{ .525,  .8,  .6}48.27\% \\
      &       & Standard Few-shot & \cellcolor[rgb]{ .784,  .906,  .824}21.48\% & \cellcolor[rgb]{ .596,  .831,  .663}40.91\% \\
      &       & Naïve CoT Learning & \cellcolor[rgb]{ .486,  .788,  .569}52.04\% & \cellcolor[rgb]{ .416,  .757,  .506}59.47\% \\
      &       & Zero-Shot VSP & \cellcolor[rgb]{ .839,  .929,  .871}15.58\% & \cellcolor[rgb]{ .529,  .804,  .608}47.56\% \\
      &       & VSP Prompting & \cellcolor[rgb]{ .467,  .78,  .553}\textbf{54.07\%} & \cellcolor[rgb]{ .404,  .753,  .494}\textbf{60.92\%} \\
      &       & Other-Type VSP & \cellcolor[rgb]{ .706,  .875,  .757}29.54\% & \cellcolor[rgb]{ .522,  .8,  .6}48.44\% \\
\cline{2-5}      & \multirow{6}[1]{*}{CVE} & Standard Prompt & \cellcolor[rgb]{ .824,  .922,  .859}17.09\% & \cellcolor[rgb]{ .647,  .851,  .706}35.39\% \\
      &       & Standard Few-shot & \cellcolor[rgb]{ .749,  .894,  .796}24.92\% & \cellcolor[rgb]{ .714,  .878,  .761}28.79\% \\
      &       & Naïve CoT Learning & \cellcolor[rgb]{ .671,  .859,  .725}33.16\% & \cellcolor[rgb]{ .612,  .835,  .675}39.17\% \\
      &       & Zero-Shot VSP & \cellcolor[rgb]{ .894,  .953,  .918}9.96\% & \cellcolor[rgb]{ .761,  .898,  .804}23.97\% \\
      &       & VSP Prompting & \cellcolor[rgb]{ .553,  .812,  .624}\textbf{45.25\%} & \cellcolor[rgb]{ .525,  .8,  .6}\textbf{48.35\%} \\
      &       & Other-Type VSP & \cellcolor[rgb]{ .902,  .953,  .925}9.20\% & \cellcolor[rgb]{ .761,  .898,  .804}23.83\% \\ \hline
\multirow{12}[2]{*}{Falcon} & \multirow{6}[1]{*}{SARD} & Standard Prompt & \cellcolor[rgb]{ .988,  .988,  1}0.17\% & \cellcolor[rgb]{ .988,  .988,  1}0.39\% \\
      &       & Standard Few-shot & \cellcolor[rgb]{ .859,  .937,  .886}13.78\% & \cellcolor[rgb]{ .871,  .941,  .898}12.46\% \\
      &       & Naïve CoT Learning & \cellcolor[rgb]{ .808,  .918,  .847}18.81\% & \cellcolor[rgb]{ .769,  .902,  .812}22.90\% \\
      &       & Zero-Shot VSP & \cellcolor[rgb]{ .984,  .988,  .996}0.63\% & \cellcolor[rgb]{ .984,  .988,  .996}0.79\% \\
      &       & VSP Prompting & \cellcolor[rgb]{ .545,  .812,  .62}\textbf{46.01\%} & \cellcolor[rgb]{ .537,  .808,  .612}\textbf{46.83\%} \\
      &       & Other-Type VSP & \cellcolor[rgb]{ .949,  .973,  .965}4.28\% & \cellcolor[rgb]{ .91,  .957,  .933}8.13\% \\
\cline{2-5}      & \multirow{6}[1]{*}{CVE} & Standard Prompt & \cellcolor[rgb]{ .984,  .988,  .996}0.69\% & \cellcolor[rgb]{ .976,  .984,  .992}1.22\% \\
      &       & Standard Few-shot & \cellcolor[rgb]{ .859,  .937,  .886}13.77\% & \cellcolor[rgb]{ .843,  .929,  .878}15.06\% \\
      &       & Naïve CoT Learning & \cellcolor[rgb]{ .871,  .941,  .898}12.54\% & \cellcolor[rgb]{ .878,  .945,  .906}11.52\% \\
      &       & Zero-Shot VSP & \cellcolor[rgb]{ .988,  .988,  1}0.00\% & \cellcolor[rgb]{ .761,  .898,  .804}23.97\% \\
      &       & VSP Prompting & \cellcolor[rgb]{ .706,  .875,  .757}\textbf{29.61\%} & \cellcolor[rgb]{ .718,  .878,  .769}\textbf{28.11\%} \\
      &       & Other-Type VSP & \cellcolor[rgb]{ .98,  .984,  .992}1.15\% & \cellcolor[rgb]{ .961,  .976,  .976}3.19\% \\ \hline
\end{tabular}%
}
    \caption{Vulnerability Discovery Results in terms of F1.}
  \label{tab:discovery-f1}
  \vspace{-0pt}
\end{table}

\begin{table}[t]
    \centering
    \scalebox{0.7}{
    \begin{tabular}{|c|c|c|c|c|}
    \hline
    &\multicolumn{2}{c|}{CVE}& \multicolumn{2}{c|}{SARD} \\ \hline
    Failure Reason& FN & FP & FN & FP \\ \hline
    Insufficient contexts& 21\% & 38\% & 2\% & 4\% \\ 
    Oblivion of CWE& 14\% & 14\% & 11\% & 4\% \\ 
    Incomplete control flow analysis& 39\% & 34\% & 57\% & 59\% \\ 
    Incomplete data flow analysis& 26\% & 21\% & 31\% & 33\%\\ 
    \hline
    \end{tabular}}
    \caption{Failure Reasons for Vulnerability Discovery}
    \label{tab:discovery-failure}
    \vspace{-0pt}
\end{table}

\subsection{Failure Case Studies}
We again do failure case studies for vulnerability discovery. To be consistent with task 1, we still do case studies on the GPT-3.5 results. We manually inspect the false negative and false positive cases on the CVE and SARD datasets separately and we inspect 100 testing samples for each case study.

By inspecting the failure cases, we still categorize the failure reasons into the same four categories. Table~\ref{tab:discovery-failure} shows the statistics. Similar to the results in task 1, we notice that the reason insufficient context is much more common on the CVE dataset, with 21\% in the false negative cases and 38\% in the false positive cases. In comparison, on the SARD dataset, the numbers are 2\% and 4\%. Again, this is caused by the complex structures of the real-world samples from the CVE dataset. 

We notice that the oblivion of CWE is still common in vulnerability discovery, with 14\% and 14\% proportions on the false negative and false positive cases against the CVE dataset, and with 11\% and 4\% proportions on the false negative and false positive cases against the SARD dataset. While the LLM needs to decide the CWE by itself in this task, it still "forgets" the CWE it is analyzing in some cases. For example, in some cases, the model decides to analyze the issues buffer write, and then in that paragraph, the model starts to analyze NULL pointer dereference. This indicates that the current LLMs still have limitations on complex and long logical analysis. 

We also notice that incomplete data flow and control flow analysis are still common in the vulnerability discovery task, with 21\%-59\% proportions, while it is more common on the SARD dataset. This again indicates that while LLMs have the capability to understand code semantics and do control flow and data flow analysis to some extent, it is still a challenge for LLMs to do complex and comprehensive code analysis.

\section{Task 3: Vulnerability Patching}
\vspace{-2pt}
\subsection{Metrics}
In this task, the LLMs are asked to generate a patch that tells what lines are modified/added/removed with the format of the {\tt diff} command outputs. However, the code style of the generated patch may be different from the ground-truth patches. Therefore, to evaluate the correctness of the generated patches, we manually check each generated patch. If the code patched by the generated patch has completely the same behavior as the one patched by the ground-truth patch (i.e., semantically equivalent), we mark the patch as correct. Therefore, we use accuracy, the proportion of correct patches in all the generated patches, as the metric to evaluate vulnerability patching. 

\begin{table*}[t]
  \centering
  \scalebox{0.7}{
\begin{tabular}{|c|c|l|r|c|c|l|r|c|c|l|r|}
\hline
Model & Dataset & \multicolumn{1}{c|}{Strategy} & \multicolumn{1}{c|}{Accuracy} & Model & Dataset & \multicolumn{1}{c|}{Strategy} & \multicolumn{1}{c|}{Accuracy} & Model & Dataset & \multicolumn{1}{c|}{Strategy} & \multicolumn{1}{c|}{Accuracy} \\
\hline
\multirow{10}[4]{*}{GPT-3.5} & \multirow{5}[2]{*}{SARD} & Standard Prompting & \cellcolor[rgb]{ .584,  .827,  .651}65.88\% & \multirow{10}[4]{*}{Llama2} & \multirow{5}[2]{*}{SARD} & Standard Prompting & \cellcolor[rgb]{ .914,  .961,  .937}12.35\% & \multirow{10}[4]{*}{Falcon} & \multirow{5}[2]{*}{SARD} & Standard Prompting & \cellcolor[rgb]{ .988,  .988,  1}0.00\% \\
      &       & Standard Few-Shot & \cellcolor[rgb]{ .396,  .749,  .49}96.47\% &       &       & Standard Few-Shot & \cellcolor[rgb]{ .937,  .969,  .957}8.82\% &       &       & Standard Few-Shot & \cellcolor[rgb]{ .988,  .988,  1}0.00\% \\
      &       & Zero-Shot VSP & \cellcolor[rgb]{ .714,  .878,  .761}45.29\% &       &       & Zero-Shot VSP & \cellcolor[rgb]{ .89,  .949,  .914}16.47\% &       &       & Zero-Shot VSP & \cellcolor[rgb]{ .988,  .988,  1}0.00\% \\
      &       & VSP Prompting & \cellcolor[rgb]{ .388,  .745,  .482}\textbf{97.65\%} &       &       & VSP Prompting & \cellcolor[rgb]{ .863,  .937,  .894}\textbf{20.59\%} &       &       & VSP Prompting & \cellcolor[rgb]{ .988,  .988,  1}0.00\% \\
      &       & Other-Type VSP & \cellcolor[rgb]{ .435,  .765,  .522}90.59\% &       &       & Other-Type VSP & \cellcolor[rgb]{ .925,  .965,  .945}10.59\% &       &       & Other-Type VSP & \cellcolor[rgb]{ .988,  .988,  1}0.00\% \\
\cline{2-4}\cline{6-8}\cline{10-12}      & \multirow{5}[2]{*}{CVE} & Standard Prompting & \cellcolor[rgb]{ .941,  .969,  .957}8.24\% &       & \multirow{5}[2]{*}{CVE} & Standard Prompting & \cellcolor[rgb]{ .976,  .984,  .988}2.35\% &       & \multirow{5}[2]{*}{CVE} & Standard Prompting & \cellcolor[rgb]{ .988,  .988,  1}0.00\% \\
      &       & Standard Few-Shot & \cellcolor[rgb]{ .898,  .953,  .922}15.29\% &       &       & Standard Few-Shot & \cellcolor[rgb]{ .91,  .957,  .933}12.94\% &       &       & Standard Few-Shot & \cellcolor[rgb]{ .988,  .988,  1}0.00\% \\
      &       & Zero-Shot VSP & \cellcolor[rgb]{ .961,  .98,  .976}4.71\% &       &       & Zero-Shot VSP & \cellcolor[rgb]{ .941,  .969,  .957}8.24\% &       &       & Zero-Shot VSP & \cellcolor[rgb]{ .988,  .988,  1}0.00\% \\
      &       & VSP Prompting & \cellcolor[rgb]{ .867,  .941,  .894}\textbf{20.00\%} &       &       & VSP Prompting & \cellcolor[rgb]{ .882,  .945,  .91}\textbf{17.65\%} &       &       & VSP Prompting & \cellcolor[rgb]{ .988,  .988,  1}0.00\% \\
      &       & Other-Type VSP & \cellcolor[rgb]{ .953,  .976,  .973}5.88\% &       &       & Other-Type VSP & \cellcolor[rgb]{ .945,  .973,  .965}7.06\% &       &       & Other-Type VSP & \cellcolor[rgb]{ .988,  .988,  1}0.00\% \\
\hline
\end{tabular}
}
     \caption{Vulnerability patching results in terms of accuracy.}
     \label{tab:patch-acc}
     \vspace{-4pt}
\end{table*}%


Since SARD and CVE only provide vulnerable and patched code, manual labeling of vulnerable locations is necessary, due to potential disparities between the locations of vulnerability and patching, as illustrated in Figure~\ref{fig:illustrate} (d). Given the time-intensive nature of this process, we restrict our evaluation to samples involving single-line edits from both SARD and CVE. This yields a final selection of 170 samples from SARD and 85 from CVE. Additionally, as LLMs possess prior knowledge of the CWE and the vulnerable location when tasked with patching, we exclude the naive CoT learning baseline from this task.

\subsection{Results}
\textbf{Overall Performance.} 
Table~\ref{tab:patch-acc} shows the effectiveness of VSP prompting and the baselines against the vulnerability patching task in terms of accuracy. The detailed accuracy of each CWE is shown in Table~\ref{tab:patching-accuracy} in Appendix. 
On the SARD dataset, GPT-3.5 and Llama2 achieve 97.65\% and 20.59\% accuracy, respectively. On the CVE dataset, GPT-3.5 and Llama2 achieve 20.00\% and 17.65\%. GPT-3.5 achieves high accuracy on the SARD dataset because the patching patterns are very simple. 
In contrast, all the LLMs achieve relatively low accuracy on CVE, because of the complexity of the real-world samples. Falcon completely fails on this task. The reason is that Falcon can process up to 2,048 tokens, which is nearly equivalent to the size of the testing samples. This limited token capacity left little room for the inclusion of the vulnerable code to patch. This indicates the necessity to use a capable LLM to do vulnerability patching, as indicated in other studies~\cite{wu2023effective,pearce2023examining}. Nevertheless, the improvements on GPT-3.5 and Llama2 from VSP prompting still indicate its effectiveness for vulnerability patching. 

\textbf{Comparison to Standard Prompting.}
As shown in Table~\ref{tab:patch-acc}, standard prompting has poor accuracy. On SARD, GPT-3.5 and Llama2 achieve 65.88\% and 12.35\% accuracy. On CVE, GPT-3.5 and Llama2 achieve 8.24\% and 2.35\% accuracy. This indicates that vulnerability patching is challenging for LLMs without prompting engineering. 

\textbf{Comparison to Standard Few-Shot Learning.}
As shown in Table~\ref{tab:patch-acc}, standard few-shot learning achieves relatively high accuracy but the accuracy is still lower than the one achieved by VSP prompting. On the SARD dataset, GPT-3.5 and Llama2 achieve 96.47\% and 8.82\% accuracy, respectively. On the CVE dataset, GPT-3.5 and Llama2 achieve 15.29\% and 12.94\% accuracy, respectively. Based on our observations on other baselines, we notice that, without few-shot learning or VSP prompting, the LLMs may output the whole patched code sample even if we explicitly say "only showing the code changes needed rather than the entire patched code". The long output may distract the LLMs and make the patch incorrect. In comparison, the exemplars in the standard few-shot learning and VSP prompting help standardize the format of the patches, thus improving the overall accuracy. This indicates the importance of exemplars for vulnerability patching.

\textbf{Comparison to Zero-Shot VSP.} To evaluate the zero-shot VSP prompting strategy, we tell the LLMs to generate patches step by step with the following prompt: "Step 1. Root cause analysis - analyze the root cause based on the given vulnerable line and the contexts with control flow and data flow analysis; Step 2. Patching Strategy - find a patching strategy which can eliminate the vulnerability based on the root cause analysis." This prompt describes the reasoning steps of VSP prompting. However, as Table~\ref{tab:patch-acc} shows, zero-shot VSP has much worse accuracy compared to VSP prompting. On the SARD dataset, GPT-3.5 and Llama2 achieve 45.29\% and 16.47\% accuracy, respectively. On the CVE dataset, GPT-3.5 and Llama2 achieve 4.71\% and 8.24\% accuracy, respectively. Because of the complexity of the reasoning steps for vulnerability patching, it is difficult for LLMs to understand the reasoning steps without exemplars. This indicates the importance of exemplars for vulnerability patching.

\textbf{Transferability of VSP Prompting.} We use exemplars that have different CWEs to the testing samples to do the evaluation. As Table~\ref{tab:patch-acc} row "Other-Type VSP" shows, the accuracy is much less than the one achieved by VSP prompting. On the SARD dataset, GPT-3.5 and Llama2 achieve 90.59\% and 10.59\% accuracy, respectively. On the CVE dataset, GPT-3.5 and Llama2 achieve 5.88\% and 7.06\% accuracy, respectively. One possible reason is that vulnerability patching is more complex than vulnerability identification and vulnerability discovery. Each CWE has different patching strategies which are difficult to transfer to another CWE.

\begin{table}[t]
    \centering
    \scalebox{0.7}{
    \begin{tabular}{|c|c|c|}
    \hline
    Failure Reason & CVE & SARD \\ \hline
    Insufficient contexts & 44.26\% & 0.00\%  \\ 
    Oblivion of CWE & 19.67\% & 0.00\%  \\ 
    Incomplete control flow analysis & 21.31\% & 0.00\% \\ 
    Incomplete data flow analysis & 14.75\% & 100.00\%\\ 
    \hline
    \end{tabular}}
    \caption{Failure Reasons for Vulnerability Patching}
    \label{tab:patching-failure}
    \vspace{-4pt}
\end{table}

\subsection{Failure Case Studies}
We again perform case studies related to instances where patching the vulnerability did not yield successful results. As Table~\ref{tab:patching-failure} shows, 
similar to observations in the previous tasks, the insufficient context caused failures are remarkably prevalent in the CVE dataset (accounting for 44.26\% of the cases), because of the complex structures of real-world projects. 


The issue of oblivion of CWE is substantial within the CVE dataset, comprising 19.67\% of cases. This issue occurs when the model, even after identifying the vulnerable lines, forgets the specific CWE it needs to address during the patching process. Again, this lapse in memory is partially due to the incapability of LLMs for long text. 

Incomplete control and data flow analysis accounts for 21.31\% and 14.75\% of the failure cases on the CVE dataset, respectively. This is caused by the model's inability to perform a comprehensive analysis for all possible execution paths. In the case of SARD, all the failure cases are related to data flow. Therefore, even with synthetic data, the model still faces some challenges in performing comprehensive data flow analysis in handling vulnerability patching.

\vspace{-0pt}
\section{Discussion}
\vspace{-0pt}

In this section, we further explore {\tech}'s potential by examining its help with discovering zero-days using LLMs. We also discuss the rationales behind {\tech}'s merits, as well as its limitations and improvement strategies. Finally, we compare {\tech} with recent peer approaches to vulnerability analysis.

\begin{figure}[t]
\centering
	\includegraphics[width=0.98\linewidth]{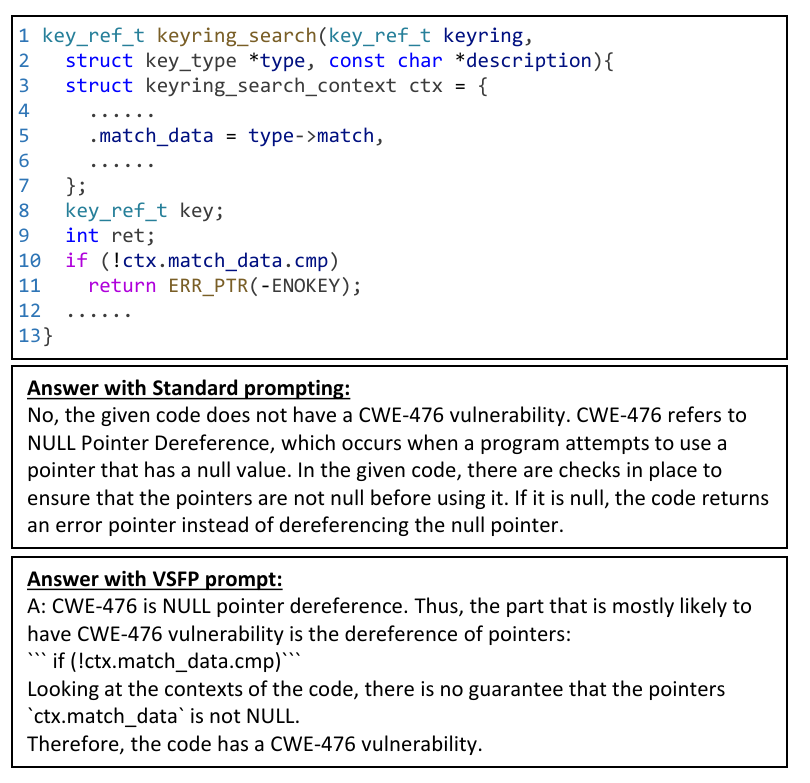}
	\caption{An example of GPT-3.5 identifying a CWE-476 vulnerability with {\tech} but not with standard prompting.}
	\label{fig:whygood}
\end{figure}

\vspace{-0pt}
\subsection{Zero-Day Vulnerability Discovery}
%
We evaluate {\tech}'s ability to detect recent zero-days, we pick a legacy LLM "GPT-3.5-Turbo-0301", frozen as of March 1st, 2023, without subsequent updates. We then collect 55 vulnerability samples from 19 critical software projects (e.g., Linux kernel) reported post-March 1st, 2023, from the CVE/NVD database~\cite{nvd}. Employing our {\tech} prompting strategy, the model correctly discovers 22 true-positive vulnerabilities at 40.00\% accuracy. Without {\tech}, only 9 vulnerabilities are detected accurately, resulting in a mere 16.36\% accuracy. Table~\ref{tab:discovery-new-full} in the Appendix shows the detailed list of these vulnerabilities. This demonstrates the potential of our {\tech} strategy in aiding LLMs to uncover zero-day vulnerabilities.

    
    
    
    
    
    
    

\subsection{Why {\tech} Works}
The main reason that {\tech} improves LLMs' vulnerability analysis effectiveness is that it makes LLMs focus on the parts that are most important for understanding vulnerabilities, which are vulnerability semantics in this work. Figure~\ref{fig:whygood} shows an example that GPT-3.5 correctly identifies a CWE-476 vulnerability with {\tech} 
but fails with standard prompting. With standard prompting, GPT-3.5 just simply scans the code and notices that there are checks to ensure the pointers are not NULL, thus it predicts the code as non-vulnerable. However, it does not attend to the pointer dereference and the contexts for further analysis. Indeed, while the check at line 10 ensures that {\tt ctx.match\_data.cmp} is not NULL, it does not ensure that {\tt ctx.match\_data} is not NULL. Without focusing on that part of the code, it is easy to ignore this vulnerability. In contrast, with {\tech}, GPT-3.5 focuses on that part of the code and notices the dereference of pointer {\tt ctx.match\_data} may be NULL. In this case, the model correctly identifies the vulnerability. This indicates that focusing on vulnerability semantics is the main reason why {\tech} works better. 

Another reason for {\tech}'s greater effectiveness is that 
it eliminates the irrelevant information that may distract the LLMs in the reasoning steps. This can be seen from our results on naive CoT learning underperforming {\tech}. 
With 
naive CoT learning, the 
code 
is analyzed line by line, no matter whether the lines are relevant to vulnerabilities, causing distractions hence diluting the core reasoning needed. 

\begin{figure}[t]
\centering
	\includegraphics[width=0.98\linewidth]{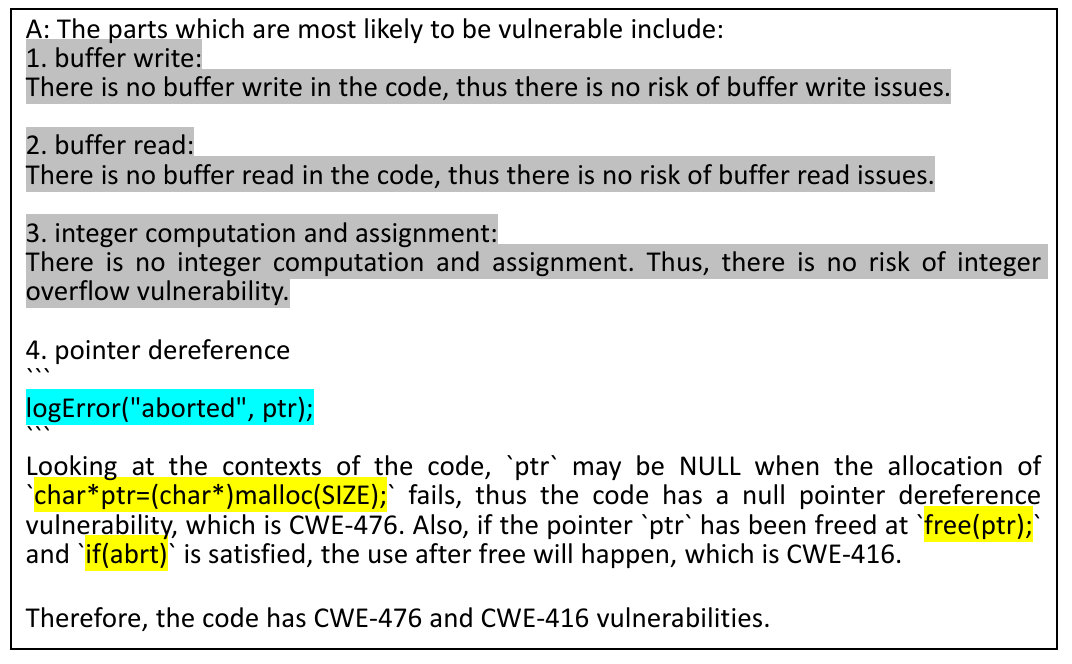}
	\caption{The exemplar corresponding to Figure~\ref{fig:illustrate} (c) involving unnecessary text (marked as gray) in the reasoning.}
	\label{fig:irrelevant}
\end{figure}


To further affirm the second reason, we conduct an experiment on vulnerability discovery. In the exemplars, we reason about not only vulnerabilities specific to the sample but also other, common types of vulnerabilities. Figure~\ref{fig:irrelevant} illustrates an instance with extraneous text related to Figure~\ref{fig:illustrate} (c). Including this unnecessary information reduces GPT-3.5's micro-averaged F1 from 36.34\% to 26.37\%, and the macro-averaged F1 from 36.89\% to 17.02\% on the CVE dataset. This underscores the importance of eliminating irrelevant details in the reasoning process for {\tech} to attain good effectiveness.

\subsection{Failure Causes and Recommendations}
\subsubsection{Insufficient Contexts} 
We notice that it is common for LLMs to fail on real-world samples because of insufficient contexts. Because real-world projects are usually large and involve many functions/files, many vulnerabilities are inter-procedural. This makes the vulnerability analysis difficult if we only provide a single function. However, given that LLMs can only process a limited number of tokens at a time (e.g., 16K for GPT-3.5), it is not likely to input the whole project into an LLM for analysis. 

\textbf{Therefore, a possible solution for this issue is to provide the context information in the code (e.g., by adding code comments).} For example, when an external function is called in the function analyzed, we can add a comment telling what the external function does and what safety guarantees are done or not done in the function. Figure~\ref{fig:contexts} in Appendix shows an example of adding comments for the external function calls. 
After doing so, the original failure is avoided. 


\subsubsection{Oblivion of CWE.} 
%
%
We notice that 
LLMs occasionally 'forget' the CWE they are analyzing, possibly due to the great length of input and output text (discussed in Section~\ref{ssec:iden-failure}). To verify this, we conducted a code length analysis for different failure reasons on the CVE dataset, as presented in Table~\ref{tab:length}. The average length for samples exhibiting 'oblivion' of CWE cases is notably higher than for others. This suggests that 'oblivion' is more likely to occur in longer code samples.

\textbf{A potential remedy for this issue is to relocate the question after the code sample and explicitly define the meaning of the CWEs.} An example of vulnerability identification is provided in Figure~\ref{fig:oblivion} in the Appendix. After doing this, the generated text is closer to the CWE ID with explicit meaning in the question, which prevents the "forgetting" issue.


\begin{table}[t]
\vspace{5pt}
    \centering
    \scalebox{0.8}{
    \begin{tabular}{|c|c|}
        \hline
        Samples & Average \#bytes\\
        \hline
        Insufficient context cases & 4,222 \\
        Oblivion of CWE cases & \textbf{5,881}\\
        Incomplete control flow analysis & 4,254\\
        Incomplete data flow analysis & 3,154\\
        \textit{All CVE samples} & \textit{3,854} \\ \hline
    \end{tabular}}
    \caption{Average \#bytes in the failure cases versus all the samples from the CVE dataset.}
    \label{tab:length}
\end{table}

\subsubsection{Incomplete Control and Data Flow Analysis}
\vspace{-0pt}
We notice that, while LLMs have the ability to understand code semantics and perform code-based analysis to some extent, they fall short of conducting comprehensive control and data flow analysis, comparable to conventional static/dynamic techniques. 
Given the complexity of real-world code, replete with various control flow (e.g., branching) structures and variable definitions/uses, relying solely on LLMs is insufficient as it could lead to exceedingly long reasoning processes. 

\textbf{Thus, for advanced vulnerability analysis, we recommend integrating traditional code analysis techniques with {\tech}-based approaches.} Specifically, static code analysis tools may be employed to extract the contexts of (potentially) vulnerable lines, as the actual vulnerability contexts may be located far from these lines.

\vspace{-0pt}
\subsection{Comparison to Recent Works}
\vspace{-0pt}
\textbf{Comparison to LLM-based Works.}
Pearce et al.~\cite{pearce2023examining} recently investigated zero-shot vulnerability repair using LLMs, conducting experiments on automatically generated, hand-crafted, and real-world datasets. They assessed various zero-shot prompting strategies and parameter configurations. However, their datasets were tailored specifically for vulnerability repair testing and are relatively straightforward. Additionally, their real-world dataset \textit{included the edited line directly in the prompt}, which is \textit{less practical}. In contrast, our study evaluates LLMs on publicly available datasets with realistic conditions, providing only the vulnerable line. Furthermore, our novel prompting strategy, focusing on vulnerability semantics with exemplars, offers a comprehensive evaluation of the potential and limitations of LLMs.

\noindent
\textbf{Comparison to Traditional DL-based Works.}
DL-based vulnerability analysis has shown promising effectiveness 
in both detection and repair~\cite{li2018vuldeepecker,zhou2019devign,chakraborty2021deep,fu2022linevul,mirskyvulchecker}. However, their performance is often overestimated due to unrealistic testing setups, such as using part of training sets for testing~\cite{nong2022open}. When evaluated on independent (unseen) real-world testing sets, these techniques yield much lower results, with detection techniques achieving up to 16.43\% F1 and repair techniques up to 8.55\% top-1 accuracy~\cite{nongvgx}. 
One way to mitigate the poor performance is data augmentation~\cite{nong2022generating,nongvulgen}. 
However, even after augmenting the training dataset with 15,039 high-quality vulnerable samples and even more normal samples using the latest data-augmentation technique, 
those numbers only went up to 20.1\% for detection and 21.05\% for repair~\cite{nongvgx}. 
In contrast, {\tech} achieves up to 48.35\% F1 for vulnerability discovery (more challenging than their binary detection), and up to 20\% top-1 accuracy for patching 
using just 20 exemplars. 
While our real-world testing dataset is different from those in~\cite{nongvgx}, {\tech} on LLMs shows better potential than existing DL-based approaches to vulnerability analysis. 

\section{Related Work}\label{sec:related}
%
\vspace{-4pt}
\noindent
{\bf Vulnerability identification/discovery.}
Traditionally, vulnerability detection 
is addressed through static and/or dynamic code analysis~\cite{li2010comparative,nong2021evaluating}. 
For instance, PCA~\cite{li2020pca} detects memory leaks via static data-flow analysis focusing on efficiency optimization, while KMeld~\cite{emamdoost2021detecting} targets the same type of vulnerabilities induced by specialized memory allocation/deallocation functions. 
CBMC~\cite{kroening2014cbmc} discovers a range of memory-error related vulnerabilities through (static) model checking, validating assertions against memory-safety violations. 

With dynamic analysis, 
undangle~\cite{caballero2012undangle} identifies use-after-free and double-free vulnerabilities via searching dangling pointers, and TT-XSS~\cite{wang2018tt} identifies DOM cross-site scripting (XSS) vulnerabilities via dynamic taint tracking. 
Both targeting buffer overruns, Lhee and Chapin~\cite{lhee2002type} focuses on 
array boundary checking, while Cred~\cite{ruwase2004practical} checks the bounds of various kinds of memory accesses. 
More broadly, IntentDroid~\cite{hay2015dynamic} detects eight types of 
inter-app communication (IAC) vulnerabilities as a result of unsafe handling of incoming IAC messages, while DrMemory~\cite{bruening2011practical} and Valgrind~\cite{nethercote2007valgrind} discover 
memory-safety vulnerabilities 
through memory shadowing. 
Similarly, AddressSanitizer~\cite{serebryany2012addresssanitizer} also discovers a variety of out-of-bounds access 
as well as use-after-free errors, while FlowDist~\cite{fu2021flowdist} and PolyCruise~\cite{li2022polycruise} detects 
various types of taint-style vulnerabilities---but these techniques work by combining static and dynamic analysis techniques. 

A main class of dynamic approach is fuzzing, which generates run-time inputs to trigger vulnerabilities~\cite{zhu2022fuzzing}. 
For instance, UAFuzz~\cite{nguyen2020binary} identifies use-after-free (UaF) vulnerabilities 
through directed greybox fuzzing, while 
UAFL~\cite{wang2020typestate} models the same type of vulnerabilities as typestate properties and checks property violations during fuzzing. 
FUZE~\cite{wu2018fuze} also focuses on detecting UaF bugs in OS kernels by combining kernel fuzzing and symbolic execution. 
Likewise, Dowser~\cite{haller2013dowser} addresses a particular kind of (i.e., buffer overflow/underflow) vulnerabilities using taint tracking combined with 
symbolic execution. 
More fuzzing approaches, however, are devised to discover a wider range of vulnerabilities~\cite{afl-fuzz,lipolyfuzz,atheris,manes2019art}.

Another major direction in vulnerability detection is to leverage  
machine learning, especially deep learning (DL)~\cite{chakraborty2021deep,nong2022open}. 
For example, VulCNN~\cite{wu2022vulcnn} detects vulnerabilities by modeling programs as images hence leveraging the merits of convolutional neural network (CNN).
Like Devign~\cite{zhou2019devign}, VulChecker~\cite{mirskyvulchecker} learns a graph neural network (GNN) based program representation for detecting vulnerabilities. 
Many other DL-based approaches, such as LineVul~\cite{fu2022linevul} and LineVD~\cite{hin2022linevd}, detect vulnerabilities through sequence modeling of code as natural language tokens based on Transformer~\cite{vaswani2017attention}. In addition,  IVDetect~\cite{li2021vulnerability} offers explainability alongside detected vulnerabilities, and like LineVul and LineVD it advances over function-level detection~\cite{li2021sysevr,chakraborty2021deep}.

Our work explores vulnerability identification/discovery using LLMs, which generally falls in the DL-based category. A key difference lies in that traditional DL-based detection relies on a sizable labeled training dataset, which LLM-prompting-based like ours only needs a few prompting exemplars. 

\vspace{2pt}
\noindent
{\bf Vulnerability patching/repair.} 
VuRLE~\cite{ma2017vurle} repairs vulnerable code by clustering code transformation edits in a set of repair examples, similar to 
Seader~\cite{zhang2022example} using fixing edit patterns.  
ExtractFix~\cite{gao2021beyond} fixes a vulnerability for which the test (exploit) is available using symbolic execution. 
VRepair~\cite{chen2022neural} attempts to fix security vulnerabilities by utilizing knowledge learned from functionality-bug fixes through transfer learning based on a vanilla Transformer~\cite{vaswani2017attention}. 
Also by fine-tuning Transformer but leveraging a pre-trained code model, VulRepair~\cite{fu2022vulrepair} improves repair accuracy over VRepair by using the T5 Transformer~\cite{wang2021codet5} 
along with the BPE tokenization~\cite{gage1994new}. 
Lately, Pearce et al.~\cite{pearce2023examining} applied several LLMs in the zero-shot setting to vulnerability repair, showing the challenges with repairing real-world vulnerabilities. 

In comparison, we explore LLM-prompting-based approaches which do not rely on (sizable) fine-tuning datasets or known exploits. 
Also, unlike~\cite{pearce2023examining}, we also examine various settings beyond zero-shot on LLMs.

\vspace{2pt}
\noindent
{\bf Prompting LLMs.}
Different from prompt learning~\cite{liu2023pre,he2023you} and fine tuning~\cite{yang2021few} as part of model \textit{training} 
(which thus requires downstream-task-specific labeled datasets),  
prompting is an \textit{inference}-time technique to improve (pre-trained) model's responses. It also does not iteratively adjust prompts as in 
prompt tuning~\cite{lester2021power}.
For instance, chain of thought (CoT) prompting elicits complex multi-step reasoning through step-by-step answer examples~\cite{wei2022chain}, 
while zero-shot CoT is a CoT variant replacing those examples with a simple prompt (``Let’s think step by step")~\cite{kojima2022large}. 
Later, tree of thoughts (ToT) generalizes over CoT by considering multiple different reasoning paths to allow for deliberate decision making~\cite{yao2023tree}, and 
graph of thoughts (GoT)~\cite{besta2023graph} further generalizes CoT by 
modeling the LLM's reasoning process as a graph, enhancing model capabilities through networked reasoning. 

In contrast, while our {\tech} approach is initially inspired by the general CoT methodology, it differs from it in focusing on selectively prompting LLMs only with the thoughts (i.e., vulnerability semantics) that are most relevant to the downstream tasks (i.e., vulnerability analysis).
\section{Conclusion}\label{sec:conclude}
We performed the first extensive study exploring how to leverage the potential of LLMs for software vulnerability analysis through vulnerability-semantics-guided prompting ({\tech}). 
Inspired by and instantiating chain-of-thought (CoT) prompting, {\tech} focuses on the most essential program behaviors that cause vulnerabilities when demonstrating to LLMs with the vulnerability analysis reasoning steps. 
Our experiments and case studies on three representative 
defense tasks against software vulnerabilities (identification, discovery, and patching) on two datasets and three LLMs reveal impressive merits of {\tech} over five LLM-based approaches as baselines.

\bibliographystyle{plain}
\bibliography{paper}

\begin{thebibliography}{10}

\bibitem{cwereport}
2022 cwe top 25 most dangerous software weaknesses.
\newblock \url{https://cwe.mitre.org/top25/archive/2022/2022_cwe_top25.html}, 2022.

\bibitem{aho2006compilers}
Alfred~V Aho, Monica~S Lam, Ravi Sethi, and Jeffrey~D Ullman.
\newblock Compilers: Principles, techniques, and tools, 2006.

\bibitem{antunes2009comparing}
Nuno Antunes and Marco Vieira.
\newblock Comparing the effectiveness of penetration testing and static code analysis on the detection of {SQL} injection vulnerabilities in web services.
\newblock In {\em Pacific Rim International Symposium on Dependable Computing}, pages 301--306, 2009.

\bibitem{aota2020automation}
Masaki Aota, Hideaki Kanehara, Masaki Kubo, Noboru Murata, Bo~Sun, and Takeshi Takahashi.
\newblock Automation of vulnerability classification from its description using machine learning.
\newblock In {\em 2020 IEEE Symposium on Computers and Communications (ISCC)}, pages 1--7. IEEE, 2020.

\bibitem{austin2013comparison}
Andrew Austin, Casper Holmgreen, and Laurie Williams.
\newblock A comparison of the efficiency and effectiveness of vulnerability discovery techniques.
\newblock {\em Information and Software Technology}, 55(7):1279--1288, 2013.

\bibitem{besta2023graph}
Maciej Besta, Nils Blach, Ales Kubicek, Robert Gerstenberger, Lukas Gianinazzi, Joanna Gajda, Tomasz Lehmann, Michal Podstawski, Hubert Niewiadomski, Piotr Nyczyk, et~al.
\newblock Graph of thoughts: Solving elaborate problems with large language models.
\newblock {\em arXiv preprint arXiv:2308.09687}, 2023.

\bibitem{black2017sard}
Paul~E Black et~al.
\newblock {SARD}: A software assurance reference dataset.
\newblock In {\em Anonymous Cybersecurity Innovation Forum.()}, 2017.

\bibitem{brown2020language}
Tom Brown, Benjamin Mann, Nick Ryder, Melanie Subbiah, Jared~D Kaplan, Prafulla Dhariwal, Arvind Neelakantan, Pranav Shyam, Girish Sastry, Amanda Askell, et~al.
\newblock Language models are few-shot learners.
\newblock {\em Advances in neural information processing systems}, 33:1877--1901, 2020.

\bibitem{bruening2011practical}
Derek Bruening and Qin Zhao.
\newblock Practical memory checking with dr. memory.
\newblock In {\em International Symposium on Code Generation and Optimization (CGO 2011)}, pages 213--223. IEEE, 2011.

\bibitem{caballero2012undangle}
Juan Caballero, Gustavo Grieco, Mark Marron, and Antonio Nappa.
\newblock Undangle: early detection of dangling pointers in use-after-free and double-free vulnerabilities.
\newblock In {\em Proceedings of the 2012 International Symposium on Software Testing and Analysis}, pages 133--143, 2012.

\bibitem{chakraborty2021deep}
Saikat Chakraborty, Rahul Krishna, Yangruibo Ding, and Baishakhi Ray.
\newblock Deep learning based vulnerability detection: Are we there yet.
\newblock {\em IEEE Transactions on Software Engineering (TSE)}, 2021.

\bibitem{chen2022neural}
Zimin Chen, Steve Kommrusch, and Martin Monperrus.
\newblock Neural transfer learning for repairing security vulnerabilities in c code.
\newblock {\em IEEE Transactions on Software Engineering}, 49(1):147--165, 2022.

\bibitem{emamdoost2021detecting}
Navid Emamdoost, Qiushi Wu, Kangjie Lu, and Stephen McCamant.
\newblock Detecting kernel memory leaks in specialized modules with ownership reasoning.
\newblock In {\em The 2021 Annual Network and Distributed System Security Symposium (NDSS'21)}, 2021.

\bibitem{vulconsequence231}
{Ericsson}.
\newblock Software vulnerability: Impact \& ways to avoid it.
\newblock \url{https://www.ericsson.com/en/security/vulnerability-management}, 2023.

\bibitem{fan2023large}
Angela Fan, Beliz Gokkaya, Mark Harman, Mitya Lyubarskiy, Shubho Sengupta, Shin Yoo, and Jie~M Zhang.
\newblock Large language models for software engineering: Survey and open problems.
\newblock {\em arXiv preprint arXiv:2310.03533}, 2023.

\bibitem{fan2020ac}
Jiahao Fan, Yi~Li, Shaohua Wang, and Tien~N Nguyen.
\newblock A c/c++ code vulnerability dataset with code changes and cve summaries.
\newblock In {\em Proceedings of the 17th International Conference on Mining Software Repositories (MSR)}, pages 508--512, 2020.

\bibitem{feng2023prompting}
Sidong Feng and Chunyang Chen.
\newblock Prompting is all your need: Automated android bug replay with large language models.
\newblock {\em arXiv preprint arXiv:2306.01987}, 2023.

\bibitem{floridi2020gpt}
Luciano Floridi and Massimo Chiriatti.
\newblock Gpt-3: Its nature, scope, limits, and consequences.
\newblock {\em Minds and Machines}, 30:681--694, 2020.

\bibitem{vulconsequence232}
{Forbes Technology Council}.
\newblock Zero-day vulnerabilities: 17 consequences and complications.
\newblock \url{https://www.forbes.com/sites/forbestechcouncil/2023/05/26/zero-day-vulnerabilities-17-consequences-and-complications/?sh=711e37204b41}, 2023.

\bibitem{fu2022linevul}
Michael Fu and Chakkrit Tantithamthavorn.
\newblock {LineVul}: a transformer-based line-level vulnerability prediction.
\newblock In {\em Proceedings of the 19th International Conference on Mining Software Repositories (MSR)}, pages 608--620, 2022.

\bibitem{fu2022vulrepair}
Michael Fu, Chakkrit Tantithamthavorn, Trung Le, Van Nguyen, and Dinh Phung.
\newblock {VulRepair}: a t5-based automated software vulnerability repair.
\newblock In {\em Proceedings of the 30th ACM Joint European Software Engineering Conference and Symposium on the Foundations of Software Engineering (ESEC/FSE)}, pages 935--947, 2022.

\bibitem{fu2021flowdist}
Xiaoqin Fu and Haipeng Cai.
\newblock {FlowDist}:multi-staged refinement-based dynamic information flow analysis for distributed software systems.
\newblock In {\em 30th USENIX Security Symposium (USENIX Security 21)}, pages 2093--2110, 2021.

\bibitem{gage1994new}
Philip Gage.
\newblock A new algorithm for data compression.
\newblock {\em C Users Journal}, 12(2):23--38, 1994.

\bibitem{gao2021beyond}
Xiang Gao, Bo~Wang, Gregory~J Duck, Ruyi Ji, Yingfei Xiong, and Abhik Roychoudhury.
\newblock Beyond tests: Program vulnerability repair via crash constraint extraction.
\newblock {\em ACM Transactions on Software Engineering and Methodology (TOSEM)}, 30(2):1--27, 2021.

\bibitem{atheris}
{google}.
\newblock {A Coverage-Guided, Native Python Fuzzer}.
\newblock \url{https://github.com/google/atheris}, 2022.

\bibitem{haller2013dowser}
Istvan Haller, Asia Slowinska, Matthias Neugschwandtner, and Herbert Bos.
\newblock Dowsing for {Overflows}: A guided fuzzer to find buffer boundary violations.
\newblock In {\em 22nd USENIX Security Symposium (USENIX Security 13)}, pages 49--64, Washington, D.C., August 2013. USENIX Association.

\bibitem{hay2015dynamic}
Roee Hay, Omer Tripp, and Marco Pistoia.
\newblock Dynamic detection of inter-application communication vulnerabilities in android.
\newblock In {\em Proceedings of the 2015 International Symposium on Software Testing and Analysis}, pages 118--128, 2015.

\bibitem{he2023controlling}
Jingxuan He and Martin Vechev.
\newblock Controlling large language models to generate secure and vulnerable code.
\newblock {\em arXiv preprint arXiv:2302.05319}, 2023.

\bibitem{he2023you}
Xinlei He, Savvas Zannettou, Yun Shen, and Yang Zhang.
\newblock You only prompt once: On the capabilities of prompt learning on large language models to tackle toxic content.
\newblock {\em arXiv preprint arXiv:2308.05596}, 2023.

\bibitem{hin2022linevd}
David Hin, Andrey Kan, Huaming Chen, and M~Ali Babar.
\newblock {LineVD}: statement-level vulnerability detection using graph neural networks.
\newblock In {\em Proceedings of the 19th International Conference on Mining Software Repositories (MSR)}, pages 596--607, 2022.

\bibitem{iannone2022secret}
Emanuele Iannone, Roberta Guadagni, Filomena Ferrucci, Andrea De~Lucia, and Fabio Palomba.
\newblock The secret life of software vulnerabilities: A large-scale empirical study.
\newblock {\em IEEE Transactions on Software Engineering}, 49(1):44--63, 2022.

\bibitem{vulcost23}
{Ilan Peleg}.
\newblock The high cost of security vulnerabilities.
\newblock \url{https://www.forbes.com/sites/forbesbusinesscouncil/2023/04/10/the-high-cost-of-security-vulnerabilities-why-observability-is-the-solution/?sh=90da08612ae6}, 2023.

\bibitem{cvedashboard23}
{Information Technology Laboratory at NIST}.
\newblock National vulnerability database (nvd) dashboard.
\newblock \url{https://nvd.nist.gov/general/nvd-dashboard}, 2023.

\bibitem{johnson2013don}
Brittany Johnson, Yoonki Song, Emerson Murphy-Hill, and Robert Bowdidge.
\newblock Why don't software developers use static analysis tools to find bugs?
\newblock In {\em 2013 35th International Conference on Software Engineering (ICSE)}, pages 672--681. IEEE, 2013.

\bibitem{joshi2023repair}
Harshit Joshi, Jos{\'e}~Cambronero Sanchez, Sumit Gulwani, Vu~Le, Gust Verbruggen, and Ivan Radi{\v{c}}ek.
\newblock Repair is nearly generation: Multilingual program repair with llms.
\newblock In {\em Proceedings of the AAAI Conference on Artificial Intelligence}, volume~37, pages 5131--5140, 2023.

\bibitem{kang2023large}
Sungmin Kang, Juyeon Yoon, and Shin Yoo.
\newblock Large language models are few-shot testers: Exploring llm-based general bug reproduction.
\newblock In {\em 2023 IEEE/ACM 45th International Conference on Software Engineering (ICSE)}, pages 2312--2323. IEEE, 2023.

\bibitem{kojima2022large}
Takeshi Kojima, Shixiang~Shane Gu, Machel Reid, Yutaka Matsuo, and Yusuke Iwasawa.
\newblock Large language models are zero-shot reasoners.
\newblock {\em Advances in neural information processing systems}, 35:22199--22213, 2022.

\bibitem{kroening2014cbmc}
Daniel Kroening and Michael Tautschnig.
\newblock {CBMC}--c bounded model checker.
\newblock In {\em International Conference on Tools and Algorithms for the Construction and Analysis of Systems (TACAS)}, pages 389--391, 2014.

\bibitem{lester2021power}
Brian Lester, Rami Al-Rfou, and Noah Constant.
\newblock The power of scale for parameter-efficient prompt tuning.
\newblock {\em arXiv preprint arXiv:2104.08691}, 2021.

\bibitem{lhee2002type}
Kyung-suk Lhee and Steve~J Chapin.
\newblock $\{$Type-Assisted$\}$ dynamic buffer overflow detection.
\newblock In {\em 11th USENIX Security Symposium (USENIX Security 02)}, 2002.

\bibitem{li2023hitchhiker}
Haonan Li, Yu~Hao, Yizhuo Zhai, and Zhiyun Qian.
\newblock The hitchhiker's guide to program analysis: A journey with large language models.
\newblock {\em arXiv preprint arXiv:2308.00245}, 2023.

\bibitem{li2010comparative}
Peng Li and Baojiang Cui.
\newblock A comparative study on software vulnerability static analysis techniques and tools.
\newblock In {\em International Conference on Information Theory and Information Security}, pages 521--524, 2010.

\bibitem{li2020pca}
Wen Li, Haipeng Cai, Yulei Sui, and David Manz.
\newblock {PCA}: memory leak detection using partial call-path analysis.
\newblock In {\em Proceedings of the 28th ACM Joint Meeting on European Software Engineering Conference and Symposium on the Foundations of Software Engineering (ESEC/FSE-Demo)}, pages 1621--1625, 2020.

\bibitem{li2022polycruise}
Wen Li, Jiang Ming, Xiapu Luo, and Haipeng Cai.
\newblock $\{$PolyCruise$\}$: A $\{$Cross-Language$\}$ dynamic information flow analysis.
\newblock In {\em 31st USENIX Security Symposium (USENIX Security 22)}, pages 2513--2530, 2022.

\bibitem{lipolyfuzz}
Wen Li, Jinyang Ruan, Guangbei Yi, Long Cheng, Xiapu Luo, and Haipeng Cai.
\newblock {PolyFuzz}: Holistic greybox fuzzing of multi-language systems.
\newblock In {\em 32nd {USENIX} Security Symposium ({USENIX Security 23})}, 2023.

\bibitem{li2021vulnerability}
Yi~Li, Shaohua Wang, and Tien~N Nguyen.
\newblock Vulnerability detection with fine-grained interpretations.
\newblock In {\em Proceedings of the 29th ACM Joint Meeting on European Software Engineering Conference and Symposium on the Foundations of Software Engineering}, pages 292--303, 2021.

\bibitem{li2021sysevr}
Zhen Li, Deqing Zou, Shouhuai Xu, Hai Jin, Yawei Zhu, and Zhaoxuan Chen.
\newblock Sysevr: A framework for using deep learning to detect software vulnerabilities.
\newblock {\em IEEE Transactions on Dependable and Secure Computing}, 2021.

\bibitem{li2018vuldeepecker}
Zhen Li, Deqing Zou, Shouhuai Xu, Xinyu Ou, Hai Jin, Sujuan Wang, Zhijun Deng, and Yuyi Zhong.
\newblock Vuldeepecker: A deep learning-based system for vulnerability detection.
\newblock In {\em Network and Distributed System Security (NDSS) Symposium}, 2018.

\bibitem{liu2023your}
Jiawei Liu, Chunqiu~Steven Xia, Yuyao Wang, and Lingming Zhang.
\newblock Is your code generated by chatgpt really correct? rigorous evaluation of large language models for code generation.
\newblock {\em arXiv preprint arXiv:2305.01210}, 2023.

\bibitem{liu2023pre}
Pengfei Liu, Weizhe Yuan, Jinlan Fu, Zhengbao Jiang, Hiroaki Hayashi, and Graham Neubig.
\newblock Pre-train, prompt, and predict: A systematic survey of prompting methods in natural language processing.
\newblock {\em ACM Computing Surveys}, 55(9):1--35, 2023.

\bibitem{ma2017vurle}
Siqi Ma, Ferdian Thung, David Lo, Cong Sun, and Robert~H Deng.
\newblock Vurle: Automatic vulnerability detection and repair by learning from examples.
\newblock In {\em Computer Security--ESORICS 2017: 22nd European Symposium on Research in Computer Security, Oslo, Norway, September 11-15, 2017, Proceedings, Part II 22}, pages 229--246. Springer, 2017.

\bibitem{manes2019art}
Valentin~JM Man{\`e}s, HyungSeok Han, Choongwoo Han, Sang~Kil Cha, Manuel Egele, Edward~J Schwartz, and Maverick Woo.
\newblock The art, science, and engineering of fuzzing: A survey.
\newblock {\em IEEE Transactions on Software Engineering}, 47(11):2312--2331, 2019.

\bibitem{mirskyvulchecker}
Yisroel Mirsky, George Macon, Michael Brown, Carter Yagemann, Matthew Pruett, Evan Downing, Sukarno Mertoguno, and Wenke Lee.
\newblock {VulChecker}: Graph-based vulnerability localization in source code.
\newblock In {\em 32nd USENIX Security Symposium (USENIX Security 23)}, pages 6557--6574, Anaheim, CA, August 2023. USENIX Association.

\bibitem{afl-fuzz}
{M.Zalewski.}
\newblock {Technical "whitepaper" for afl-fuzz}.
\newblock \url{https://lcamtuf.coredump.cx/afl/technical_details.txt}, 2014.

\bibitem{nvd}
{National Institute of Standards and Technology (NIST)}.
\newblock {National Vulnerability Database (NVD)}.
\newblock \url{https://nvd.nist.gov}, 2022.

\bibitem{nethercote2007valgrind}
Nicholas Nethercote and Julian Seward.
\newblock Valgrind: a framework for heavyweight dynamic binary instrumentation.
\newblock {\em ACM Sigplan notices}, 42(6):89--100, 2007.

\bibitem{nguyen2020binary}
Manh-Dung Nguyen, S{\'e}bastien Bardin, Richard Bonichon, Roland Groz, and Matthieu Lemerre.
\newblock Binary-level directed fuzzing for $\{$Use-After-Free$\}$ vulnerabilities.
\newblock In {\em 23rd International Symposium on Research in Attacks, Intrusions and Defenses (RAID 2020)}, pages 47--62, 2020.

\bibitem{noever2023can}
David Noever.
\newblock Can large language models find and fix vulnerable software?
\newblock {\em arXiv preprint arXiv:2308.10345}, 2023.

\bibitem{nong2020preliminary}
Yu~Nong and Haipeng Cai.
\newblock A preliminary study on open-source memory vulnerability detectors.
\newblock In {\em 2020 IEEE 27th International Conference on Software Analysis, Evolution and Reengineering (SANER)}, pages 557--561. IEEE, 2020.

\bibitem{nong2021evaluating}
Yu~Nong, Haipeng Cai, Pengfei Ye, Li~Li, and Feng Chen.
\newblock Evaluating and comparing memory error vulnerability detectors.
\newblock {\em Information and Software Technology}, 137:106614, 2021.

\bibitem{nongvgx}
Yu~Nong, Richard Fang, Guangbei Yi, Kunsong Zhao, Xiapu Luo, Feng Chen, and Haipeng Cai.
\newblock {VGX}: Large-scale sample generation for boosting learning-based software vulnerability analyses.
\newblock In {\em IEEE/ACM International Conference on Software Engineering (ICSE)}, 2024.

\bibitem{nong2022generating}
Yu~Nong, Yuzhe Ou, Michael Pradel, Feng Chen, and Haipeng Cai.
\newblock Generating realistic vulnerabilities via neural code editing: an empirical study.
\newblock In {\em Proceedings of the 30th ACM Joint European Software Engineering Conference and Symposium on the Foundations of Software Engineering}, pages 1097--1109, 2022.

\bibitem{nongvulgen}
Yu~Nong, Yuzhe Ou, Michael Pradel, Feng Chen, and Haipeng Cai.
\newblock {VulGen}: Realistic vulnerable sample generation via pattern mining and deep learning.
\newblock In {\em IEEE/ACM International Conference on Software Engineering (ICSE)}, pages 2527--2539, 2023.

\bibitem{nong2022open}
Yu~Nong, Rainy Sharma, Abdelwahab Hamou-Lhadj, Xiapu Luo, and Haipeng Cai.
\newblock Open science in software engineering: A study on deep learning-based vulnerability detection.
\newblock {\em IEEE Transactions on Software Engineering (TSE)}, 2022.

\bibitem{pearce2023examining}
Hammond Pearce, Benjamin Tan, Baleegh Ahmad, Ramesh Karri, and Brendan Dolan-Gavitt.
\newblock Examining zero-shot vulnerability repair with large language models.
\newblock In {\em 2023 IEEE Symposium on Security and Privacy (SP)}, pages 2339--2356. IEEE, 2023.

\bibitem{mssilentpatch23}
Andre Protas and Steve Manzuik.
\newblock Skeletons in microsoft's closet - silently fixed vulnerabilities.
\newblock \url{https://www.blackhat.com/presentations/bh-europe-06/bh-eu-06-Manzuik.pdf}, 2023.

\bibitem{ruwase2004practical}
Olatunji Ruwase and Monica~S Lam.
\newblock A practical dynamic buffer overflow detector.
\newblock In {\em NDSS}, volume 2004, pages 159--169, 2004.

\bibitem{serebryany2012addresssanitizer}
Konstantin Serebryany, Derek Bruening, Alexander Potapenko, and Dmitriy Vyukov.
\newblock $\{$AddressSanitizer$\}$: A fast address sanity checker.
\newblock In {\em 2012 USENIX annual technical conference (USENIX ATC 12)}, pages 309--318, 2012.

\bibitem{touvron2023llama}
Hugo Touvron, Louis Martin, Kevin Stone, Peter Albert, Amjad Almahairi, Yasmine Babaei, Nikolay Bashlykov, Soumya Batra, Prajjwal Bhargava, Shruti Bhosale, et~al.
\newblock Llama 2: Open foundation and fine-tuned chat models.
\newblock {\em arXiv preprint arXiv:2307.09288}, 2023.

\bibitem{vaswani2017attention}
Ashish Vaswani, Noam Shazeer, Niki Parmar, Jakob Uszkoreit, Llion Jones, Aidan~N Gomez, {\L}ukasz Kaiser, and Illia Polosukhin.
\newblock Attention is all you need.
\newblock {\em Advances in Neural Information Processing Systems (NeurIPS)}, 30, 2017.

\bibitem{wang2020typestate}
Haijun Wang, Xiaofei Xie, Yi~Li, Cheng Wen, Yuekang Li, Yang Liu, Shengchao Qin, Hongxu Chen, and Yulei Sui.
\newblock Typestate-guided fuzzer for discovering use-after-free vulnerabilities.
\newblock In {\em Proceedings of the ACM/IEEE 42nd International Conference on Software Engineering}, pages 999--1010, 2020.

\bibitem{wang2023software}
Junjie Wang, Yuchao Huang, Chunyang Chen, Zhe Liu, Song Wang, and Qing Wang.
\newblock Software testing with large language model: Survey, landscape, and vision.
\newblock {\em arXiv preprint arXiv:2307.07221}, 2023.

\bibitem{wang2018tt}
Ran Wang, Guangquan Xu, Xianjiao Zeng, Xiaohong Li, and Zhiyong Feng.
\newblock Tt-xss: A novel taint tracking based dynamic detection framework for dom cross-site scripting.
\newblock {\em Journal of Parallel and Distributed Computing}, 118:100--106, 2018.

\bibitem{wang2021codet5}
Yue Wang, Weishi Wang, Shafiq Joty, and Steven~CH Hoi.
\newblock {CodeT5}: Identifier-aware unified pre-trained encoder-decoder models for code understanding and generation.
\newblock {\em arXiv preprint arXiv:2109.00859}, 2021.

\bibitem{wei2022chain}
Jason Wei, Xuezhi Wang, Dale Schuurmans, Maarten Bosma, Fei Xia, Ed~Chi, Quoc~V Le, Denny Zhou, et~al.
\newblock Chain-of-thought prompting elicits reasoning in large language models.
\newblock {\em Advances in Neural Information Processing Systems}, 35:24824--24837, 2022.

\bibitem{wei2023copiloting}
Yuxiang Wei, Chunqiu~Steven Xia, and Lingming Zhang.
\newblock Copiloting the copilots: Fusing large language models with completion engines for automated program repair.
\newblock {\em arXiv preprint arXiv:2309.00608}, 2023.

\bibitem{white2023prompt}
Jules White, Quchen Fu, Sam Hays, Michael Sandborn, Carlos Olea, Henry Gilbert, Ashraf Elnashar, Jesse Spencer-Smith, and Douglas~C Schmidt.
\newblock A prompt pattern catalog to enhance prompt engineering with chatgpt.
\newblock {\em arXiv preprint arXiv:2302.11382}, 2023.

\bibitem{wolf2019huggingface}
Thomas Wolf, Lysandre Debut, Victor Sanh, Julien Chaumond, Clement Delangue, Anthony Moi, Pierric Cistac, Tim Rault, R{\'e}mi Louf, Morgan Funtowicz, et~al.
\newblock Huggingface's transformers: State-of-the-art natural language processing.
\newblock {\em arXiv preprint arXiv:1910.03771}, 2019.

\bibitem{wu2018fuze}
Wei Wu, Yueqi Chen, Jun Xu, Xinyu Xing, Xiaorui Gong, and Wei Zou.
\newblock $\{$FUZE$\}$: Towards facilitating exploit generation for kernel $\{$Use-After-Free$\}$ vulnerabilities.
\newblock In {\em 27th USENIX Security Symposium (USENIX Security 18)}, pages 781--797, 2018.

\bibitem{wu2023effective}
Yi~Wu, Nan Jiang, Hung~Viet Pham, Thibaud Lutellier, Jordan Davis, Lin Tan, Petr Babkin, and Sameena Shah.
\newblock How effective are neural networks for fixing security vulnerabilities.
\newblock {\em arXiv preprint arXiv:2305.18607}, 2023.

\bibitem{wu2022vulcnn}
Yueming Wu, Deqing Zou, Shihan Dou, Wei Yang, Duo Xu, and Hai Jin.
\newblock Vulcnn: An image-inspired scalable vulnerability detection system.
\newblock In {\em Proceedings of the 44th International Conference on Software Engineering}, pages 2365--2376, 2022.

\bibitem{xia2023universal}
Chunqiu~Steven Xia, Matteo Paltenghi, Jia~Le Tian, Michael Pradel, and Lingming Zhang.
\newblock Universal fuzzing via large language models.
\newblock {\em arXiv preprint arXiv:2308.04748}, 2023.

\bibitem{yang2021few}
Guanqun Yang, Shay Dineen, Zhipeng Lin, and Xueqing Liu.
\newblock Few-sample named entity recognition for security vulnerability reports by fine-tuning pre-trained language models.
\newblock In {\em Deployable Machine Learning for Security Defense: Second International Workshop, MLHat 2021, Virtual Event, August 15, 2021, Proceedings 2}, pages 55--78. Springer, 2021.

\bibitem{yao2023tree}
Shunyu Yao, Dian Yu, Jeffrey Zhao, Izhak Shafran, Thomas~L Griffiths, Yuan Cao, and Karthik Narasimhan.
\newblock Tree of thoughts: Deliberate problem solving with large language models.
\newblock {\em arXiv preprint arXiv:2305.10601}, 2023.

\bibitem{zhang2022example}
Ying Zhang, Ya~Xiao, Md~Mahir~Asef Kabir, Danfeng Yao, and Na~Meng.
\newblock Example-based vulnerability detection and repair in java code.
\newblock In {\em Proceedings of the 30th IEEE/ACM International Conference on Program Comprehension}, pages 190--201, 2022.

\bibitem{zhou2019devign}
Yaqin Zhou, Shangqing Liu, Jingkai Siow, Xiaoning Du, and Yang Liu.
\newblock Devign: Effective vulnerability identification by learning comprehensive program semantics via graph neural networks.
\newblock {\em Advances in Neural Information Processing Systems (NeurIPS)}, 32, 2019.

\bibitem{zhu2022fuzzing}
Xiaogang Zhu, Sheng Wen, Seyit Camtepe, and Yang Xiang.
\newblock Fuzzing: a survey for roadmap.
\newblock {\em ACM Computing Surveys (CSUR)}, 54(11s):1--36, 2022.

\bibitem{zxhang2023falcon}
Yoshua~X ZXhang, Yann~M Haxo, and Ying~X Mat.
\newblock Falcon llm: A new frontier in natural language processing.
\newblock {\em AC Investment Research Journal}, 220(44), 2023.

\end{thebibliography}



\appendices
\section{Full Results for each Task}
\textbf{Task 1.} Tables \ref{tab:iden-recall-full}, \ref{tab:iden-precision-full}, and \ref{tab:iden-f1-full} show the recall, precision, and F1 of each CWE for Task 1.

\noindent
\textbf{Task 2.} Tables~\ref{tab:discovery-recall}, \ref{tab:discovery-precision}, and \ref{tab:discovery-f1-full} show the recall, precision, and F1 of each CWE for Task 2.

\noindent
\textbf{Task 3.} Table~\ref{tab:patching-accuracy} show the accuracy of each CWE for Task 3.

\section{Zero-day vulnerability discovery}
The full list of the latest CVEs that VSP prompting and standard prompting can and cannot discover is shown in Table~\ref{tab:discovery-new-full}.

\section{Failure Reasons and Recommendations}
\begin{figure*}[tp]
\centering
	\includegraphics[width=0.8\linewidth]{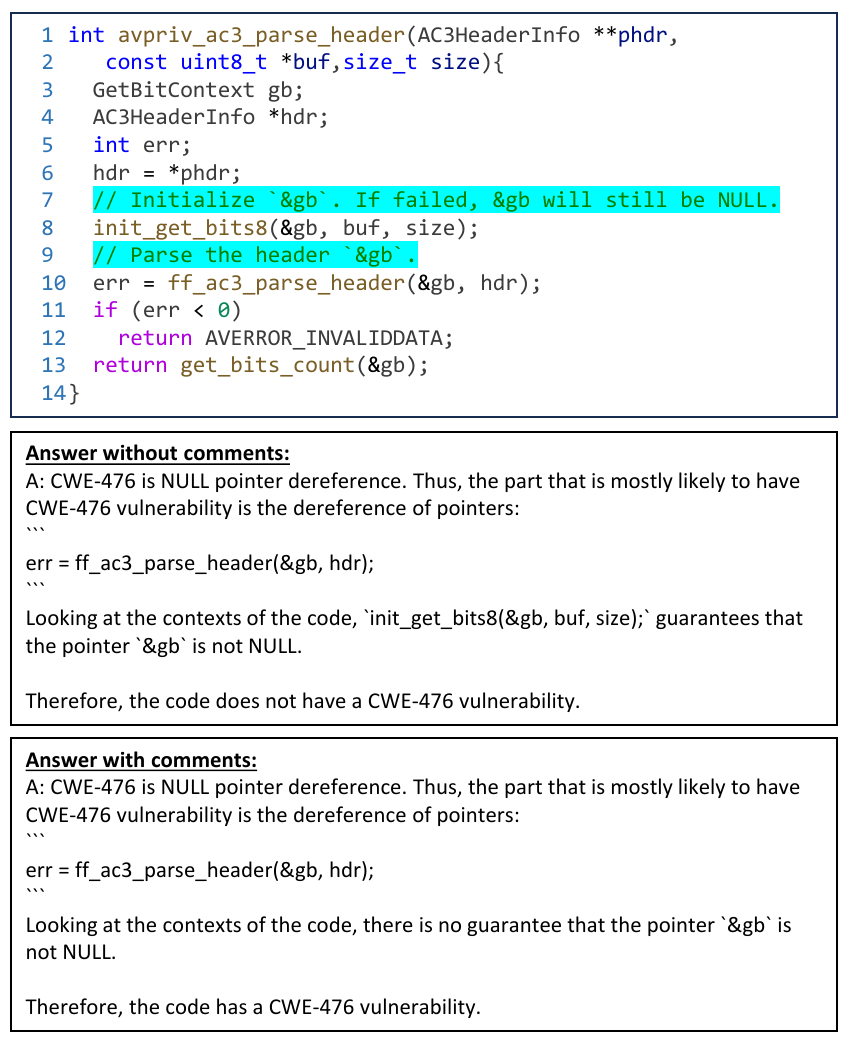}
	\caption{An example that telling the model the contexts by adding comments helps the model do correct analysis.}
	\label{fig:contexts}
\end{figure*}

\begin{figure*}[tp]
\centering
	\includegraphics[width=0.8\linewidth]{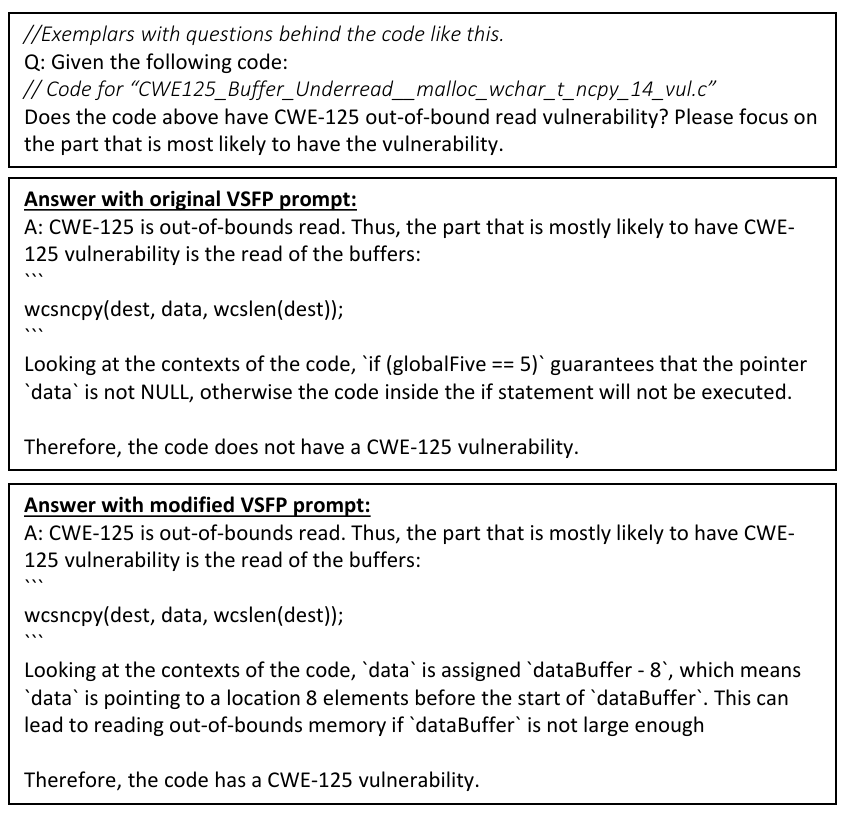}
	\caption{An example that moving the question behind the code and telling the CWE ID meaning help prevent oblivion of CWE.}
	\label{fig:oblivion}
\end{figure*}

Figures~\ref{fig:contexts} and \ref{fig:oblivion} show the examples for the failure reasons "insufficient contexts" and "oblivion of CWE" as well as their possible solutions.

\begin{table*}[tp]
  \centering
  \scalebox{0.85}{
\begin{tabular}{|c|c|l|r|r|r|r|r|r|r|}
\hline
Model & Dataset & \multicolumn{1}{c|}{Strategy} & \multicolumn{1}{c|}{CWE-787} & \multicolumn{1}{c|}{CWE-125} & \multicolumn{1}{c|}{CWE-416} & \multicolumn{1}{c|}{CWE-476} & \multicolumn{1}{c|}{CWE-190} & \multicolumn{1}{c|}{Macro avg} & \multicolumn{1}{c|}{Micro avg} \\
\hline
\multirow{12}[0]{*}{GPT-3.5} & \multirow{6}[0]{*}{SARD} & Standard Prompting & \cellcolor[rgb]{ .58,  .824,  .647}68.56\% & \cellcolor[rgb]{ .561,  .816,  .631}71.54\% & \cellcolor[rgb]{ .929,  .965,  .949}10.00\% & \cellcolor[rgb]{ .541,  .808,  .612}75.00\% & \cellcolor[rgb]{ .984,  .988,  .996}0.77\% & \cellcolor[rgb]{ .718,  .878,  .769}45.17\% & \cellcolor[rgb]{ .686,  .867,  .741}50.60\% \\
      &       & Standard Few-Shot & \cellcolor[rgb]{ .796,  .91,  .835}32.31\% & \cellcolor[rgb]{ .78,  .906,  .82}34.96\% & \cellcolor[rgb]{ .871,  .941,  .898}20.00\% & \cellcolor[rgb]{ .765,  .898,  .808}37.50\% & \cellcolor[rgb]{ .973,  .984,  .984}3.08\% & \cellcolor[rgb]{ .835,  .929,  .871}25.57\% & \cellcolor[rgb]{ .839,  .929,  .871}25.20\% \\
      &       & Naive CoT Learning & \cellcolor[rgb]{ .918,  .961,  .937}12.23\% & \cellcolor[rgb]{ .937,  .969,  .957}8.94\% & \cellcolor[rgb]{ .812,  .918,  .847}30.00\% & \cellcolor[rgb]{ .914,  .961,  .937}12.50\% & \cellcolor[rgb]{ .988,  .988,  1}0.00\% & \cellcolor[rgb]{ .914,  .961,  .937}12.73\% & \cellcolor[rgb]{ .937,  .969,  .957}8.62\% \\
      &       & Zero-Shot VSP & \cellcolor[rgb]{ .486,  .788,  .569}83.84\% & \cellcolor[rgb]{ .498,  .792,  .576}82.11\% & \cellcolor[rgb]{ .749,  .894,  .796}40.00\% & \cellcolor[rgb]{ .765,  .898,  .808}37.50\% & \cellcolor[rgb]{ .969,  .98,  .98}3.85\% & \cellcolor[rgb]{ .694,  .871,  .745}49.46\% & \cellcolor[rgb]{ .624,  .843,  .686}60.92\% \\
\hline
\end{tabular}%
}
    \caption{Full vulnerability identification effectiveness in terms of recall.}
   \label{tab:iden-recall-full}
\end{table*}%

\begin{table*}[tp]
  \centering
  \scalebox{0.85}{
\begin{tabular}{|c|c|l|r|r|r|r|r|r|r|}
\hline
Model & Dataset & \multicolumn{1}{c|}{Strategy} & \multicolumn{1}{c|}{CWE-787} & \multicolumn{1}{c|}{CWE-125} & \multicolumn{1}{c|}{CWE-416} & \multicolumn{1}{c|}{CWE-476} & \multicolumn{1}{c|}{CWE-190} & \multicolumn{1}{c|}{Macro avg} & \multicolumn{1}{c|}{Micro avg} \\
\hline
\multirow{12}[0]{*}{GPT-3.5} & \multirow{6}[0]{*}{SARD} & Standard Prompting & \cellcolor[rgb]{ .627,  .843,  .69}60.15\% & \cellcolor[rgb]{ .58,  .824,  .647}68.22\% & \cellcolor[rgb]{ .388,  .745,  .482}100.00\% & \cellcolor[rgb]{ .388,  .745,  .482}100.00\% & \cellcolor[rgb]{ .69,  .867,  .741}50.00\% & \cellcolor[rgb]{ .537,  .808,  .612}75.67\% & \cellcolor[rgb]{ .608,  .835,  .675}63.41\% \\
      &       & Standard Few-Shot & \cellcolor[rgb]{ .686,  .867,  .741}50.68\% & \cellcolor[rgb]{ .678,  .863,  .733}51.81\% & \cellcolor[rgb]{ .592,  .827,  .659}66.67\% & \cellcolor[rgb]{ .388,  .745,  .482}100.00\% & \cellcolor[rgb]{ .592,  .827,  .659}66.67\% & \cellcolor[rgb]{ .588,  .827,  .655}67.17\% & \cellcolor[rgb]{ .678,  .863,  .729}52.28\% \\
      &       & Naive CoT Learning & \cellcolor[rgb]{ .647,  .851,  .706}57.14\% & \cellcolor[rgb]{ .549,  .812,  .624}73.33\% & \cellcolor[rgb]{ .388,  .745,  .482}100.00\% & \cellcolor[rgb]{ .388,  .745,  .482}100.00\% & \cellcolor[rgb]{ .988,  .988,  1}0.00\% & \cellcolor[rgb]{ .592,  .831,  .659}66.10\% & \cellcolor[rgb]{ .612,  .835,  .675}63.24\% \\
      &       & Zero-Shot VSP & \cellcolor[rgb]{ .671,  .859,  .725}53.48\% & \cellcolor[rgb]{ .69,  .867,  .741}50.00\% & \cellcolor[rgb]{ .388,  .745,  .482}100.00\% & \cellcolor[rgb]{ .388,  .745,  .482}100.00\% & \cellcolor[rgb]{ .867,  .941,  .894}20.83\% & \cellcolor[rgb]{ .6,  .831,  .667}64.86\% & \cellcolor[rgb]{ .682,  .867,  .737}51.44\% \\
\hline
\end{tabular}%
}
       \caption{Full vulnerability identification effectiveness in terms of precision.}
   \label{tab:iden-precision-full}
\end{table*}

\begin{table*}[tp]
  \centering
  \scalebox{0.85}{
    
    \begin{tabular}{|c|c|l|r|r|r|r|r|r|r|}
    \hline
    Model & Dataset & \multicolumn{1}{c|}{Strategy} & \multicolumn{1}{c|}{CWE-787} & \multicolumn{1}{c|}{CWE-125} & \multicolumn{1}{c|}{CWE-416} & \multicolumn{1}{c|}{CWE-476} & \multicolumn{1}{c|}{CWE-190} & \multicolumn{1}{c|}{Macro avg} & \multicolumn{1}{c|}{Micro avg} \\
    \hline
    \multirow{12}[4]{*}{GPT-3.5} & \multirow{6}[2]{*}{SARD} & Standard Prompting & \cellcolor[rgb]{ .541,  .808,  .616}64.08\% & \cellcolor[rgb]{ .502,  .792,  .58}69.84\% & \cellcolor[rgb]{ .863,  .937,  .894}18.18\% & \cellcolor[rgb]{ .388,  .745,  .482}85.71\% & \cellcolor[rgb]{ .98,  .984,  .992}1.52\% & \cellcolor[rgb]{ .655,  .855,  .714}47.87\% & \cellcolor[rgb]{ .596,  .831,  .663}56.28\% \\
          &       & Standard Few-Shot & \cellcolor[rgb]{ .714,  .878,  .765}39.47\% & \cellcolor[rgb]{ .698,  .871,  .749}41.75\% & \cellcolor[rgb]{ .776,  .902,  .816}30.77\% & \cellcolor[rgb]{ .608,  .835,  .675}54.55\% & \cellcolor[rgb]{ .949,  .973,  .965}5.88\% & \cellcolor[rgb]{ .749,  .894,  .792}34.48\% & \cellcolor[rgb]{ .753,  .894,  .796}34.01\% \\
          &       & Naive CoT Learning & \cellcolor[rgb]{ .851,  .933,  .878}20.14\% & \cellcolor[rgb]{ .878,  .945,  .906}15.94\% & \cellcolor[rgb]{ .667,  .859,  .722}46.15\% & \cellcolor[rgb]{ .835,  .925,  .867}22.22\% & \cellcolor[rgb]{ .988,  .988,  1}0.00\% & \cellcolor[rgb]{ .843,  .929,  .875}20.89\% & \cellcolor[rgb]{ .882,  .949,  .91}15.17\% \\
          &       & Zero-Shot VSP & \cellcolor[rgb]{ .533,  .804,  .608}65.31\% & \cellcolor[rgb]{ .557,  .816,  .627}62.15\% & \cellcolor[rgb]{ .592,  .827,  .659}57.14\% & \cellcolor[rgb]{ .608,  .835,  .675}54.55\% & \cellcolor[rgb]{ .945,  .973,  .965}6.49\% & \cellcolor[rgb]{ .647,  .851,  .706}49.13\% & \cellcolor[rgb]{ .6,  .831,  .667}55.78\% \\
\hline
    \end{tabular}}
      \caption{Full vulnerability identification effectiveness in terms of F1.}
   \label{tab:iden-f1-full}
\end{table*}%

\begin{table*}[tp]
  \centering
  \scalebox{0.85}{
\begin{tabular}{|c|c|l|rrrrrrr|}
\hline
\multicolumn{1}{|l|}{Model} & \multicolumn{1}{l|}{Dataset} & Strategy & \multicolumn{1}{l}{CWE-787} & \multicolumn{1}{l}{CWE-125} & \multicolumn{1}{l}{CWE-476} & \multicolumn{1}{l}{CWE-416} & \multicolumn{1}{l}{CWE-190} & \multicolumn{1}{l}{Macro avg} & \multicolumn{1}{l|}{Micro avg} \\
\hline
\multirow{12}[3]{*}{GPT-3.5} & \multirow{6}[2]{*}{SARD} & Standard Prompt & \cellcolor[rgb]{ .463,  .776,  .549}83.84\% & \cellcolor[rgb]{ .816,  .922,  .851}27.64\% & \cellcolor[rgb]{ .522,  .8,  .596}75.00\% & \cellcolor[rgb]{ .616,  .839,  .678}60.00\% & \cellcolor[rgb]{ .506,  .796,  .584}76.92\% & \cellcolor[rgb]{ .671,  .863,  .725}50.84\% & \cellcolor[rgb]{ .678,  .863,  .733}49.60\% \\
      &       & Standard Few-shot & \cellcolor[rgb]{ .592,  .827,  .659}63.59\% & \cellcolor[rgb]{ .765,  .898,  .808}35.77\% & \cellcolor[rgb]{ .443,  .769,  .529}87.50\% & \cellcolor[rgb]{ .427,  .761,  .514}90.00\% & \cellcolor[rgb]{ .875,  .945,  .902}18.46\% & \cellcolor[rgb]{ .62,  .839,  .682}59.06\% & \cellcolor[rgb]{ .702,  .875,  .753}45.89\% \\
      &       & Naïve CoT Learning & \cellcolor[rgb]{ .722,  .882,  .769}42.79\% & \cellcolor[rgb]{ .894,  .949,  .918}15.45\% & \cellcolor[rgb]{ .522,  .8,  .596}75.00\% & \cellcolor[rgb]{ .804,  .914,  .839}30.00\% & \cellcolor[rgb]{ .976,  .984,  .988}2.31\% & \cellcolor[rgb]{ .784,  .906,  .824}33.11\% & \cellcolor[rgb]{ .827,  .925,  .863}25.80\% \\
      &       & Zero-Shot VSP & \cellcolor[rgb]{ .482,  .784,  .565}80.80\% & \cellcolor[rgb]{ .859,  .937,  .886}21.21\% & \cellcolor[rgb]{ .533,  .804,  .608}72.73\% & \cellcolor[rgb]{ .706,  .875,  .757}45.45\% & \cellcolor[rgb]{ .973,  .984,  .984}2.94\% & \cellcolor[rgb]{ .71,  .878,  .761}44.63\% & \cellcolor[rgb]{ .702,  .875,  .753}45.74\% \\
\hline
\end{tabular}%
}
  \caption{Full vulnerability discovery effectiveness in terms of recall.}
   \label{tab:discovery-recall}%
\end{table*}%

\begin{table*}[tp]
  \centering
    \scalebox{0.85}{
\begin{tabular}{|c|c|l|rrrrrrr|}
\hline
\multicolumn{1}{|l|}{Model} & \multicolumn{1}{l|}{Dataset} & Strategy & \multicolumn{1}{l}{CWE-787} & \multicolumn{1}{l}{CWE-125} & \multicolumn{1}{l}{CWE-476} & \multicolumn{1}{l}{CWE-416} & \multicolumn{1}{l}{CWE-190} & \multicolumn{1}{l}{Macro avg} & \multicolumn{1}{l|}{Micro avg} \\
\hline
\multirow{12}[3]{*}{GPT-3.5} & \multirow{6}[2]{*}{SARD} & Standard Prompt & \cellcolor[rgb]{ .675,  .863,  .729}52.75\% & \cellcolor[rgb]{ .631,  .847,  .694}59.65\% & \cellcolor[rgb]{ .475,  .78,  .557}85.71\% & \cellcolor[rgb]{ .388,  .745,  .482}100.00\% & \cellcolor[rgb]{ .784,  .906,  .824}34.48\% & \cellcolor[rgb]{ .592,  .827,  .659}66.52\% & \cellcolor[rgb]{ .671,  .859,  .725}53.56\% \\
      &       & Standard Few-shot & \cellcolor[rgb]{ .682,  .867,  .737}51.42\% & \cellcolor[rgb]{ .663,  .859,  .722}54.32\% & \cellcolor[rgb]{ .522,  .8,  .6}77.78\% & \cellcolor[rgb]{ .631,  .843,  .69}60.00\% & \cellcolor[rgb]{ .835,  .925,  .867}26.08\% & \cellcolor[rgb]{ .667,  .859,  .722}53.92\% & \cellcolor[rgb]{ .702,  .875,  .753}47.81\% \\
      &       & Naïve CoT Learning & \cellcolor[rgb]{ .702,  .875,  .753}48.28\% & \cellcolor[rgb]{ .655,  .855,  .714}55.88\% & \cellcolor[rgb]{ .388,  .745,  .482}100.00\% & \cellcolor[rgb]{ .388,  .745,  .482}100.00\% & \cellcolor[rgb]{ .827,  .925,  .863}27.27\% & \cellcolor[rgb]{ .592,  .827,  .659}66.28\% & \cellcolor[rgb]{ .69,  .867,  .741}50.19\% \\
      &       & Zero-Shot VSP & \cellcolor[rgb]{ .678,  .863,  .733}51.80\% & \cellcolor[rgb]{ .761,  .898,  .804}38.36\% & \cellcolor[rgb]{ .388,  .745,  .482}100.00\% & \cellcolor[rgb]{ .388,  .745,  .482}100.00\% & \cellcolor[rgb]{ .831,  .925,  .863}26.67\% & \cellcolor[rgb]{ .612,  .835,  .675}63.36\% & \cellcolor[rgb]{ .69,  .867,  .741}50.30\% \\
\hline
\end{tabular}%
}
    \caption{Full vulnerability discovery effectiveness in terms of precision.}
  \label{tab:discovery-precision}
\end{table*}

\begin{table*}[tp]
  \centering
  \scalebox{0.85}{
\begin{tabular}{|c|c|lrrrrrrr|}\hline
\multicolumn{1}{|l|}{Model} & \multicolumn{1}{l|}{Dataset} & \multicolumn{1}{l|}{Strategy} & \multicolumn{1}{l}{CWE-787} & \multicolumn{1}{l}{CWE-125} & \multicolumn{1}{l}{CWE-476} & \multicolumn{1}{l}{CWE-416} & \multicolumn{1}{l}{CWE-190} & \multicolumn{1}{l}{Macro avg} & \multicolumn{1}{l|}{Micro avg} \\
\cline{1-3}\cline{9-10}\multirow{12}[3]{*}{GPT-3.5} & \multirow{6}[2]{*}{SARD} & Standard Prompt & \cellcolor[rgb]{ .537,  .808,  .612}64.75\% & \cellcolor[rgb]{ .725,  .882,  .773}37.78\% & \cellcolor[rgb]{ .431,  .765,  .518}80.00\% & \cellcolor[rgb]{ .459,  .776,  .541}75.99\% & \cellcolor[rgb]{ .902,  .953,  .925}12.57\% & \cellcolor[rgb]{ .612,  .835,  .675}54.02\% & \cellcolor[rgb]{ .631,  .843,  .69}51.50\% \\
      &       & Standard Few-shot & \cellcolor[rgb]{ .541,  .808,  .616}64.26\% & \cellcolor[rgb]{ .596,  .831,  .663}56.14\% & \cellcolor[rgb]{ .549,  .812,  .62}63.16\% & \cellcolor[rgb]{ .729,  .882,  .776}37.50\% & \cellcolor[rgb]{ .584,  .824,  .651}58.12\% & \cellcolor[rgb]{ .604,  .835,  .671}55.19\% & \cellcolor[rgb]{ .663,  .859,  .718}46.83\% \\
      &       & Naïve CoT Learning & \cellcolor[rgb]{ .549,  .812,  .62}63.12\% & \cellcolor[rgb]{ .8,  .914,  .835}27.31\% & \cellcolor[rgb]{ .4,  .753,  .494}84.21\% & \cellcolor[rgb]{ .553,  .812,  .624}62.49\% & \cellcolor[rgb]{ .953,  .976,  .969}5.29\% & \cellcolor[rgb]{ .702,  .875,  .753}41.13\% & \cellcolor[rgb]{ .753,  .894,  .796}34.08\% \\
      &       & Zero-Shot VSP & \cellcolor[rgb]{ .675,  .863,  .729}45.37\% & \cellcolor[rgb]{ .82,  .922,  .855}24.20\% & \cellcolor[rgb]{ .388,  .745,  .482}85.71\% & \cellcolor[rgb]{ .667,  .859,  .722}46.15\% & \cellcolor[rgb]{ .961,  .976,  .976}4.26\% & \cellcolor[rgb]{ .651,  .851,  .71}48.49\% & \cellcolor[rgb]{ .655,  .855,  .714}47.91\% \\
\hline
\end{tabular}}
  \caption{Full vulnerability discovery effectiveness in terms of F1.}
   \label{tab:discovery-f1-full}
\end{table*}%

\begin{table*}[tp]
  \centering
  \scalebox{0.85}{
\begin{tabular}{|c|c|l|r|r|r|r|r|r|r|}
\hline
Model & Dataset & \multicolumn{1}{c|}{Strategy} & \multicolumn{1}{c|}{CWE-787} & \multicolumn{1}{c|}{CWE-125} & \multicolumn{1}{c|}{CWE-416} & \multicolumn{1}{c|}{CWE-476} & \multicolumn{1}{c|}{CWE-190} & \multicolumn{1}{c|}{Macro avg} & \multicolumn{1}{c|}{Micro avg} \\
\hline
\multirow{10}[4]{*}{GPT-3.5} & \multirow{5}[2]{*}{SARD} & Standard Prompting & \cellcolor[rgb]{ .604,  .835,  .671}64.29\% & \cellcolor[rgb]{ .561,  .816,  .631}71.43\% & \cellcolor[rgb]{ .988,  .988,  1}0.00\% & N/A   & N/A   & \cellcolor[rgb]{ .718,  .878,  .769}45.24\% & \cellcolor[rgb]{ .596,  .831,  .663}65.88\% \\
      &       & Standard Few-Shot & \cellcolor[rgb]{ .4,  .753,  .494}98.21\% & \cellcolor[rgb]{ .412,  .757,  .502}96.43\% & \cellcolor[rgb]{ .988,  .988,  1}0.00\% & N/A   & N/A   & \cellcolor[rgb]{ .6,  .831,  .667}64.88\% & \cellcolor[rgb]{ .412,  .757,  .502}96.47\% \\
      &       & Zero-Shot VSP & \cellcolor[rgb]{ .706,  .875,  .757}47.32\% & \cellcolor[rgb]{ .733,  .886,  .78}42.86\% & \cellcolor[rgb]{ .988,  .988,  1}0.00\% & N/A   & N/A   & \cellcolor[rgb]{ .812,  .918,  .847}30.06\% & \cellcolor[rgb]{ .718,  .878,  .769}45.29\% \\
      &       & VSP Prompting & \cellcolor[rgb]{ .4,  .753,  .494}98.21\% & \cellcolor[rgb]{ .412,  .757,  .502}96.43\% & \cellcolor[rgb]{ .388,  .745,  .482}100.00\% & N/A   & N/A   & \cellcolor[rgb]{ .4,  .753,  .494}98.21\% & \cellcolor[rgb]{ .404,  .753,  .498}97.65\% \\
      &       & Other-Type VSP & \cellcolor[rgb]{ .431,  .765,  .522}92.86\% & \cellcolor[rgb]{ .455,  .773,  .541}89.29\% & \cellcolor[rgb]{ .988,  .988,  1}0.00\% & N/A   & N/A   & \cellcolor[rgb]{ .627,  .843,  .686}60.71\% & \cellcolor[rgb]{ .447,  .769,  .533}90.59\% \\
\hline
\end{tabular}%
}
    \caption{Full vulnerability patching effectiveness in terms of accuracy.}
  \label{tab:patching-accuracy}
\end{table*}

\begin{table*}[tp]
  \centering
  \resizebox{0.6\linewidth}{!}{
    \begin{tabular}{|l|l|l|l|l|}
    \hline
    CWE ID & Project & CVE ID & VSP Prompting & Standard Prompting \\
    \hline
    CWE787 & linux & 2023-3090 & failed & failed \\
    \hline
    CWE787 & libtiff & 2023-30775 & failed & failed \\
    \hline
    CWE787 & libjpeg & 2023-2804 & failed & failed \\
    \hline
    CWE787 & gpac  & 2023-4754 & failed & failed \\
    \hline
    CWE787 & nettle & 2023-36660 & correct & failed \\
    \hline
    CWE787 & linux & 2023-35788 & correct & failed \\
    \hline
    CWE787 & radare & 2023-4322 & failed & correct \\
\hline
    \end{tabular}%
    }
    \caption{Latest vulnerabilities discovered by VSP prompting and standard prompting.}
  \label{tab:discovery-new-full}
\end{table*}

\end{document}